\documentclass[conference,10pt]{IEEEtran}
\IEEEoverridecommandlockouts

\usepackage{cite}
\usepackage{amsmath,amssymb,amsfonts}
\usepackage{algorithm}
\usepackage{algorithmic}
\usepackage{graphicx}
\usepackage{textcomp}
\usepackage{xcolor}
\usepackage{cleveref}
\usepackage{float}
\usepackage{enumitem}
\usepackage{color}
\usepackage[margin=13mm]{geometry}
\usepackage[lofdepth,lotdepth]{subfig}
\usepackage[font=footnotesize,labelfont=bf]{caption}
\usepackage{epstopdf}
\def\BibTeX{{\rm B\kern-.05em{\sc i\kern-.025em b}\kern-.08em
    T\kern-.1667em\lower.7ex\hbox{E}\kern-.125emX}}

\renewcommand{\baselinestretch}{1.0}

\definecolor{darkgreen}{RGB}{30, 160, 80}

\makeatletter

\def\ps@IEEEtitlepagestyle{%
    \def\@oddhead{Accepted for publication at Design, Automation and Test in Europe (DATE 2019). Florence, Italy \hfill}%
    \def\@evenfoot{}%
}
\def\mycopyrightnotice{%
    {\footnotesize  978-1-4799-6773-5/14/\$31.00 \textcopyright2017 Crown\hfill}
    \gdef\mycopyrightnotice{}
}



\makeatletter
\newcommand*\titleheader[1]{\gdef\@titleheader{#1}}
\AtBeginDocument{%
  \let\st@red@title\@title%
  \def\@title{%
    \bgroup\normalfont\large\centering\@titleheader\par\egroup
    \vskip1.5em\st@red@title}
}
\makeatother

\begin{document}

\author{\IEEEauthorblockN{Alberto Marchisio, Muhammad Abdullah Hanif, and Muhammad Shafique}
\IEEEauthorblockA{\textit{Vienna University of Technology, Vienna, Austria}\\
\{alberto.marchisio,muhammad.hanif,muhammad.shafique\}@tuwien.ac.at}
}



\title{CapsAcc: An Efficient Hardware Accelerator for CapsuleNets with Data Reuse}
\maketitle

\begin{abstract}

Deep Neural Networks (DNNs) have been widely deployed for many Machine Learning applications. Recently, {\em CapsuleNets} have overtaken traditional DNNs, because of their improved generalization ability due to the multi-dimensional capsules, in contrast to the single-dimensional neurons. Consequently, CapsuleNets also require extremely intense matrix computations, making it a gigantic challenge to achieve high performance. In this paper, we propose {\em CapsAcc}, the first specialized CMOS-based hardware architecture to perform CapsuleNets inference with high performance and energy efficiency. State-of-the-art convolutional DNN accelerators would not work efficiently for CapsuleNets, as their designs do not account for key operations involved in CapsuleNets, like squashing and dynamic routing, as well as multi-dimensional matrix processing. Our CapsAcc architecture targets this problem and achieves significant improvements, when compared to an optimized GPU implementation. Our architecture exploits the massive parallelism by flexibly feeding the data to a specialized systolic array according to the operations required in different layers. It also avoids extensive load and store operations on the on-chip memory, by reusing the data when possible. We further optimize the routing algorithm to reduce the computations needed at this stage. We synthesized the complete CapsAcc architecture in a $32$nm CMOS technology using Synopsys design tools, and evaluated it for the MNIST benchmark (as also done by the original CapsuleNet paper) to ensure consistent and fair comparisons. This work enables highly-efficient CapsuleNets inference on embedded platforms.

\end{abstract}


\vspace*{-2mm}
\section{Introduction}
Machine Learning (ML) algorithms are widely used for Internet of Things and Artificial Intelligence applications, such as computer vision~\cite{ref:AlexNet}, speech recognition~\cite{ref:speech_recognition} and natural language processing~\cite{ref:natural_language}. 
Deep Neural Networks (DNNs) have reached state-of-the-art results in terms of accuracy, compared to other ML algorithms. Recently, Sabour and Hinton et al.~\cite{ref:dyn_routing} proposed the Dynamic Routing algorithm to efficiently perform training and inference on CapsuleNets~\cite{ref:trans_autoencoder}. Such CapsuleNets are able to encapsulate multi-dimensional features across the layers, while traditional Convolutional Neural Networks (CNNs) do not. Thus, {\em CapsuleNets can beat traditional CNNs in multiple tasks}, like image classification, as shown in~\cite{ref:dyn_routing}. The most evident difference is that the CapsuleNets are deeper in width than in height, when compared to DNNs, as each capsule incorporates the information hierarchically, thus preserving other features like position, orientation and scaling (see an overview of CapsuleNets in~\Cref{sec:background}). The data is propagated towards the output using the so-called routing-by-agreement algorithm.

Current state-of-the-art DNN accelerators~\cite{ref:EIE} \cite{ref:Eyeriss} \cite{ref:TPU} \cite{ref:SCNN} \cite{ref:FlexFlow} proposed energy-aware solutions for inference using traditional CNNs. 
As for our knowledge, {\bf we are the first to propose a hardware accelerator-based architecture for the complete CapsuleNets inference}. Although systolic array based designs like~\cite{ref:FlexFlow} perform parallel matrix multiply-and-accumulate (MAC) operations with good efficiency, the existing CNN accelerators cannot compute several key operations of the CapsuleNets (i.e., squashing and routing-by-agreement) with high performance. An efficient data-flow mapping requires a direct feedback connection from the outputs coming from the activation unit back to the inputs of the processing element. Thus, such key optimizations can highly increase the performance and reduce the memory accesses.

\textbf{Our Novel Contributions:}
\begin{enumerate}
	\item  We analyze the memory requirements and the performance in the forward pass of CapsuleNets, through experiments on a high-end GPU, which allows to identify the corresponding bottlenecks.
	\item  We propose CapsAcc, an accelerator that can perform inference on CapsuleNets with an efficient data reuse based mapping policy.
	\item  We optimize the routing-by-agreement process at algorithm level, by skipping the first step and directly initializing the coupling coefficients.
	\item  We implement and synthesize the complete CapsAcc architecture for a $32$nm technology using the ASIC design flow, and perform evaluations for performance, area and power consumption. We performed the functional and timing validation through gate-level simulations. Our results demonstrate a speed-up of $12$$\times$ in the ClassCaps layer, of $172$$\times$ in the Squashing and of $6$$\times$ in the overall CapsuleNet inference, compared to a highly optimized GPU implementation.
\end{enumerate}

\textbf{Paper Organization:} \Cref{sec:background} summarizes the fundamental theory behind CapsuleNets and highlights the differences with traditional DNNs. In \Cref{sec:analysis}, we systematically analyze the forward pass of the CapsuleNets executing on a GPU, to identify the potential bottlenecks. \Cref{sec:design} describes the architectural design of our CapsuleNet accelerator, for which the data-flow mapping is presented in \Cref{sec:mapping}. The results are presented in \Cref{sec:evaluation}.

\section{Background: An Overview of CapsuleNets}
\label{sec:background}

Sabour and Hinton et al.~\cite{ref:dyn_routing} introduced many novelties compared to CNNs, such as the concept of capsules, the squashing activation function, and the routing-by-agreement algorithm. In this paper, since we analyze the inference process, the layers and the algorithms that are involved in the training process {\em only} (e.g., decoder, margin loss and reconstruction loss) are not discussed.

\subsection{CapsuleNet Architecture}
\label{subsec:capsnet_architecture}

\begin{figure}[t]
	\centering
	\includegraphics[width=.85\linewidth]{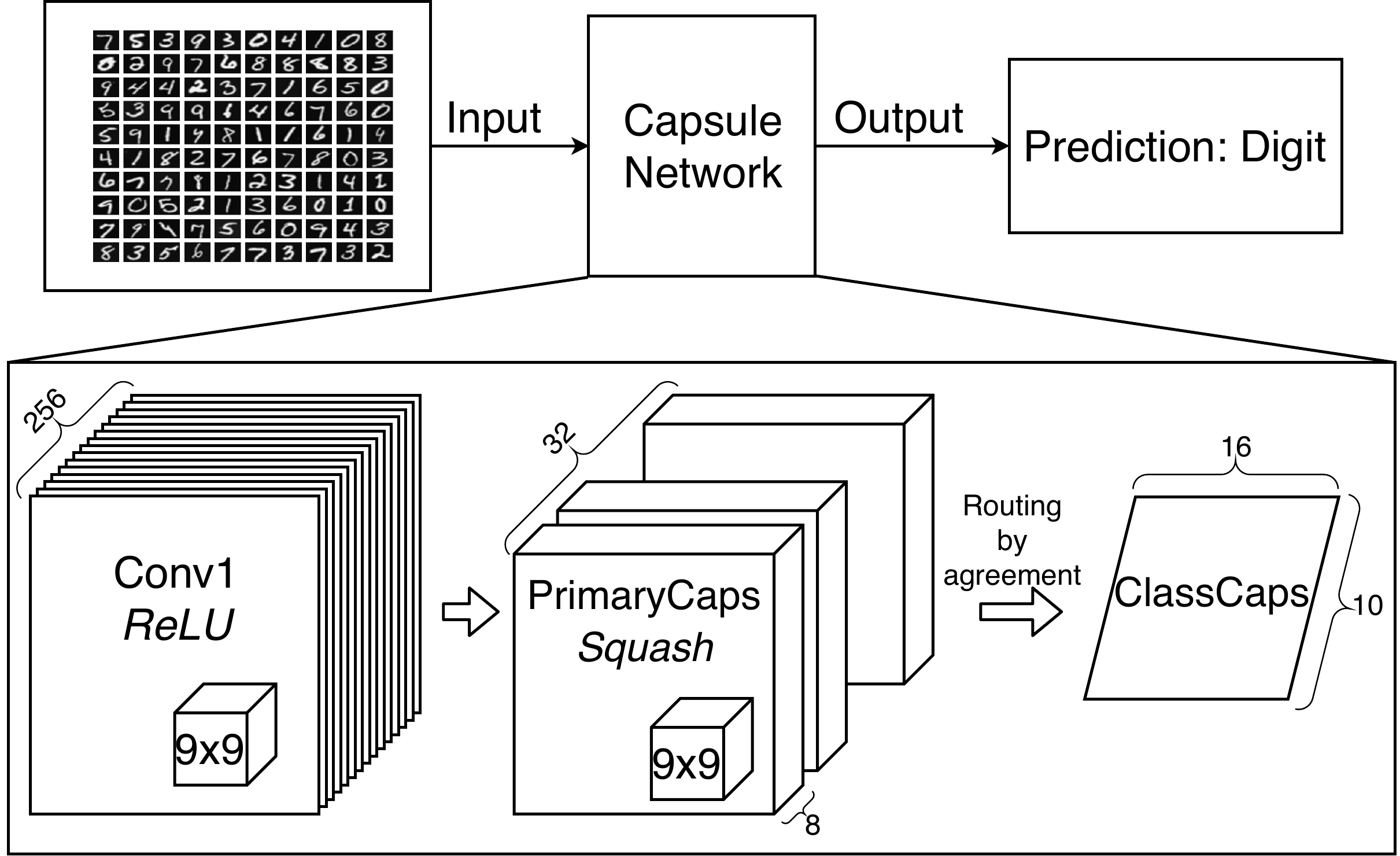}
	\caption{An overview of the CapsuleNet architecture, based on the design of \cite{ref:dyn_routing} for the MNIST dataset.}
	\label{fig:capsulenet}
    \vspace*{0mm}
\end{figure}

\Cref{fig:capsulenet} illustrates the CapsuleNet architecture~\cite{ref:dyn_routing} designed for the MNIST~\cite{ref:MNIST} dataset. 
It consists of 3 layers:
\begin{itemize}
	\item \textbf{Conv1:} traditional convolutional layer, with 256 channels, with filter size of 9x9, stride=1, and ReLU activations.
	\item \textbf{PrimaryCaps:} first capsule layer, with 32 channels. Each eight-dimensional (8D) capsule has 9x9 convolutional filters with stride=2.
	\item \textbf{ClassCaps:} last capsule layer, with 16D capsules for each output class.
\end{itemize}

\begin{figure}[t]
	\centering
	\includegraphics[width=.7\linewidth]{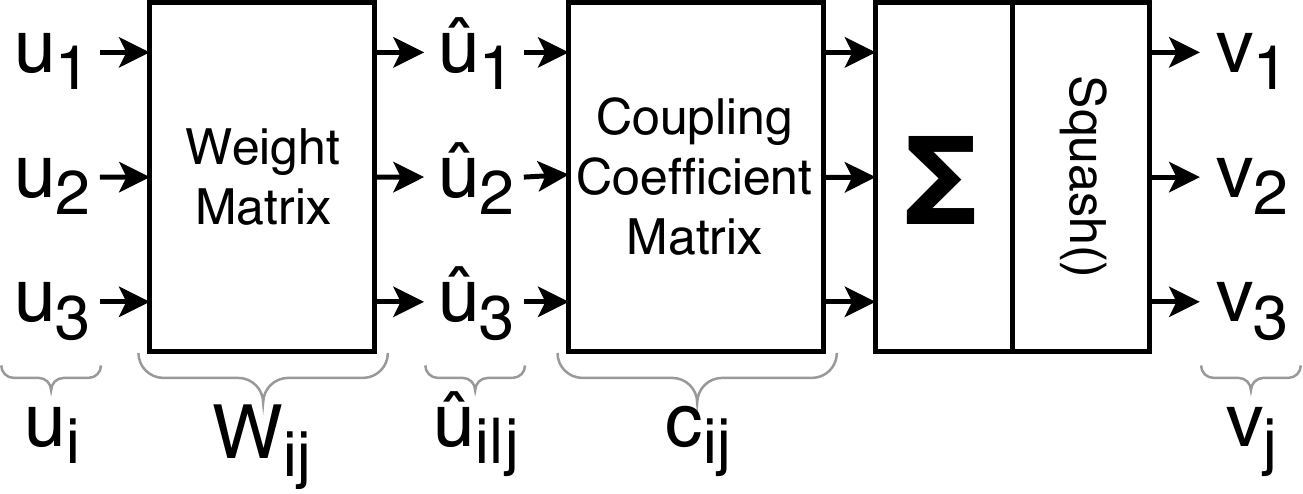}
	\caption{Simple representation of how a CapsuleNet works.}
	\label{fig:capsule_work}
    \vspace*{0mm}
\end{figure}

At the first glance, it is evident that a capsule layer contains multi-dimensional capsules, which are groups of neurons nested inside a layer. One of the main advantages of CapsuleNets over traditional CNNs is the ability to learn the hierarchy between layers, because \textit{each capsule element is able to learn different types of information} (e.g., position, orientation and scaling). Indeed, CNNs have limited model capabilities, which they try to compensate by increasing the amount of training data (with more samples and/or data augmentation) and by applying pooling to select the most important information that will be propagated to the following layers. In capsule layers, however, the outputs are propagated towards the following layers in form of a prediction vector, whose size is defined by the capsule dimension. A simple visualization of how a CapsuleNet works is presented in \Cref{fig:capsule_work}. After the weight matrix multiplication $W_{ij}$, the values $\hat{u}_{i|j}$ are multiplied by the coupling coefficients $c_{ij}$, before summing together the contributions and applying the squash function. The coupling coefficients are computed and updated at run-time during each inference pass, using the \textit{routing-by-agreement} algorithm (\Cref{fig:routing}).

\subsection{Squashing}
\label{subsec:squashing}

\begin{figure}[t]
\vspace*{0mm}
\noindent
\begin{minipage}[t]{.54\linewidth}
\begin{figure}[H]
	\centering
	\includegraphics[width=\linewidth]{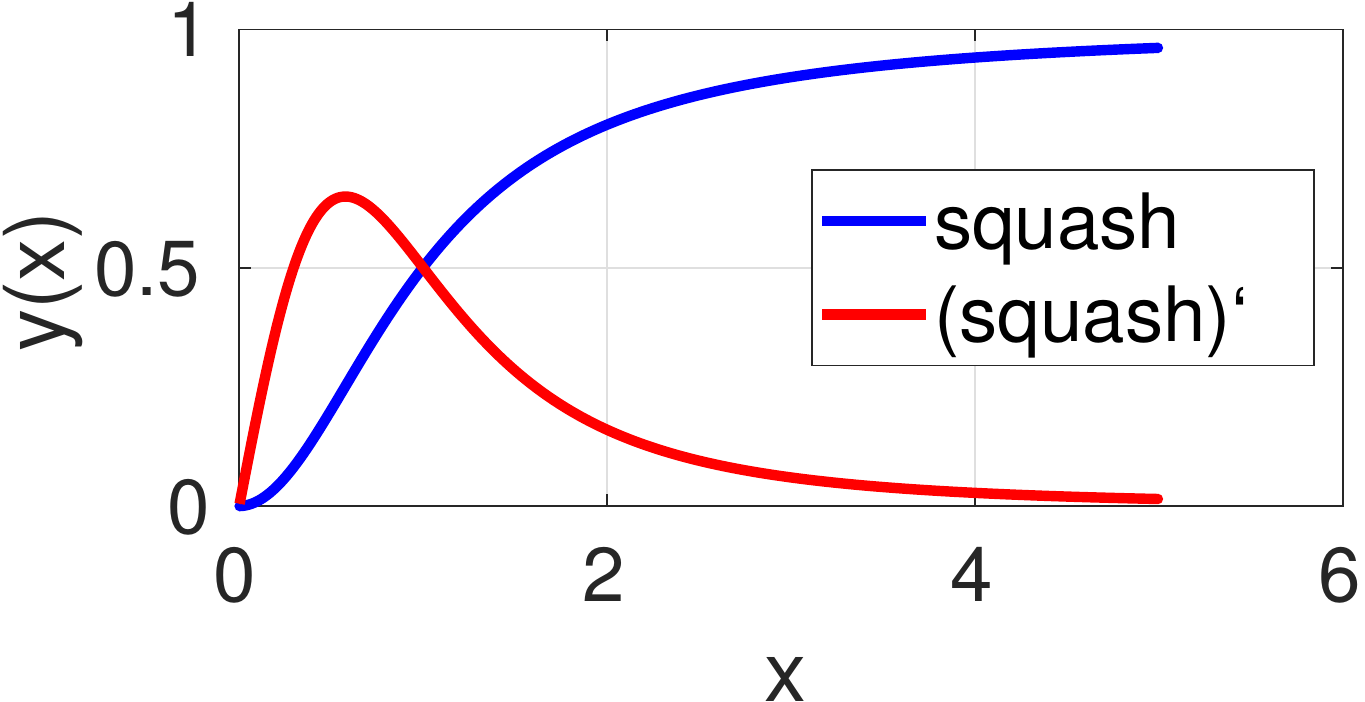}
	\caption{Squashing function and its first derivative, considering single-dimensional input.}
	\label{fig:squash_single}
\end{figure}
\end{minipage}
\hfill
\begin{minipage}[t]{.44\linewidth}
\vspace{15mm}
\begin{equation}
	v_j = \frac{\left | \left | s_j \right | \right |^2}{1 + \left | \left | s_j \right | \right |^2} \frac{s_j}{ \left | \left | s_j \right | \right |}
	\label{eq:squash}
\end{equation}
\end{minipage}
\vspace*{0mm}
\end{figure}

The squashing is an activation function designed to efficiently fit for the prediction vector. It introduces the nonlinearity into an array and normalizes the outputs to values between $0$ and $1$. Given $s_j$ as the input of the capsule $j$ (or, from another perspective, the sum of the weighted prediction vector) and $v_j$ as its respective output, the squashing function is defined by the \Cref{eq:squash}. 

The behaviors of the squashing function and its first derivative are shown in \Cref{fig:squash_single}. Note that we have plotted the single-dimensional input function, since a multi-dimensional input version cannot be visualized in a chart. The squashing function produces an output bounded between $0$ and $1$, while its first derivative follows the behavior of the red line, with a peak at the point $(0.5767, 0.6495)$.

\subsection{Routing-by-Agreement Algorithm}
\label{subsec:routing_algorithm}

\begin{figure}[t]
	\centering
	\includegraphics[width=.8\linewidth]{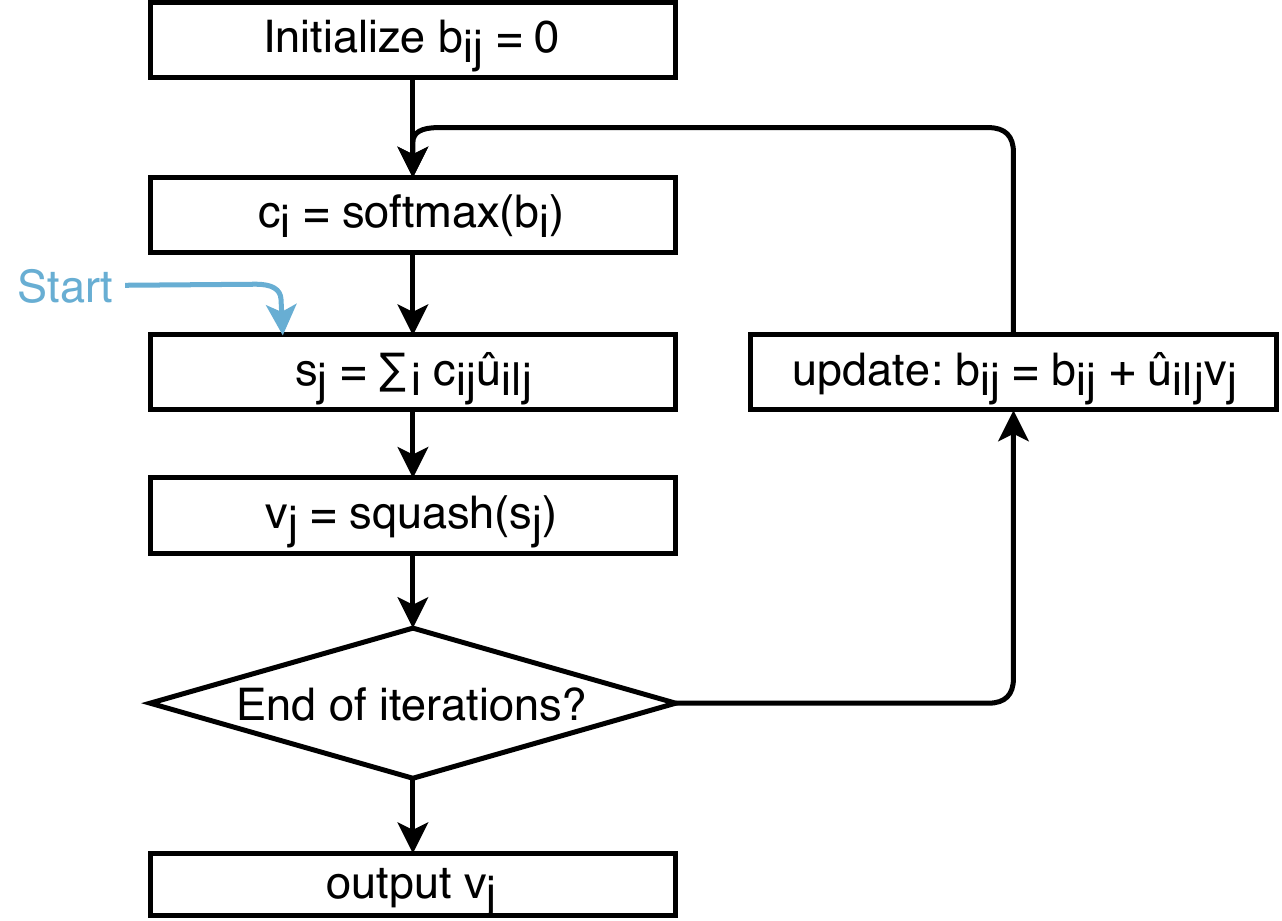}
	\caption{Flow of the routing-by-agreement algorithm.}
	\label{fig:routing}
    \vspace*{0mm}
\end{figure}

The predictions are propagated across two consecutive capsule layers through the routing-by-agreement algorithm. It is an iterative process, that introduces a feedback path in the inference pass. For clarity, we present the flow diagram (\Cref{fig:routing}) of the routing-by-agreement at software level. Note, this algorithm introduces a loop in the forward pass, because the coupling coefficients $c_{ij}$ are learned during the routing, as their values depend on the current data. Thus, they cannot be considered as constant parameters, learned during the training process. Intuitively, this step can cause a computational bottleneck, as demonstrated in \Cref{sec:analysis}.

\section{Motivational Analysis of CapsuleNet Complexity}
\label{sec:analysis}

In the following, we perform a comprehensive analysis to identify how CapsuleNet inference is performed on a standard GPU platform, like the one used in our experiments, i.e., the Nvidia Ge-Force GTX1070 GPU (see \Cref{fig:GPU_board}). First, in \Cref{subsec:trainable_parameters} we quantitatively analyze how many trainable parameters per layer must be fed from the memory. Then, in \Cref{subsec:performance_GPU} we benchmark our pyTorch~\cite{ref:pytorch} based CapsuleNet implementation for the MNIST dataset to measure the performance of the inference process on our GPU.

\subsection{Trainable parameters of the CapsuleNet}
\label{subsec:trainable_parameters}

\begin{figure}[t]
\vspace*{0mm}
\noindent
\begin{minipage}[t]{.4\linewidth}
\begin{figure}[H]
	\centering
	\includegraphics[width=\linewidth]{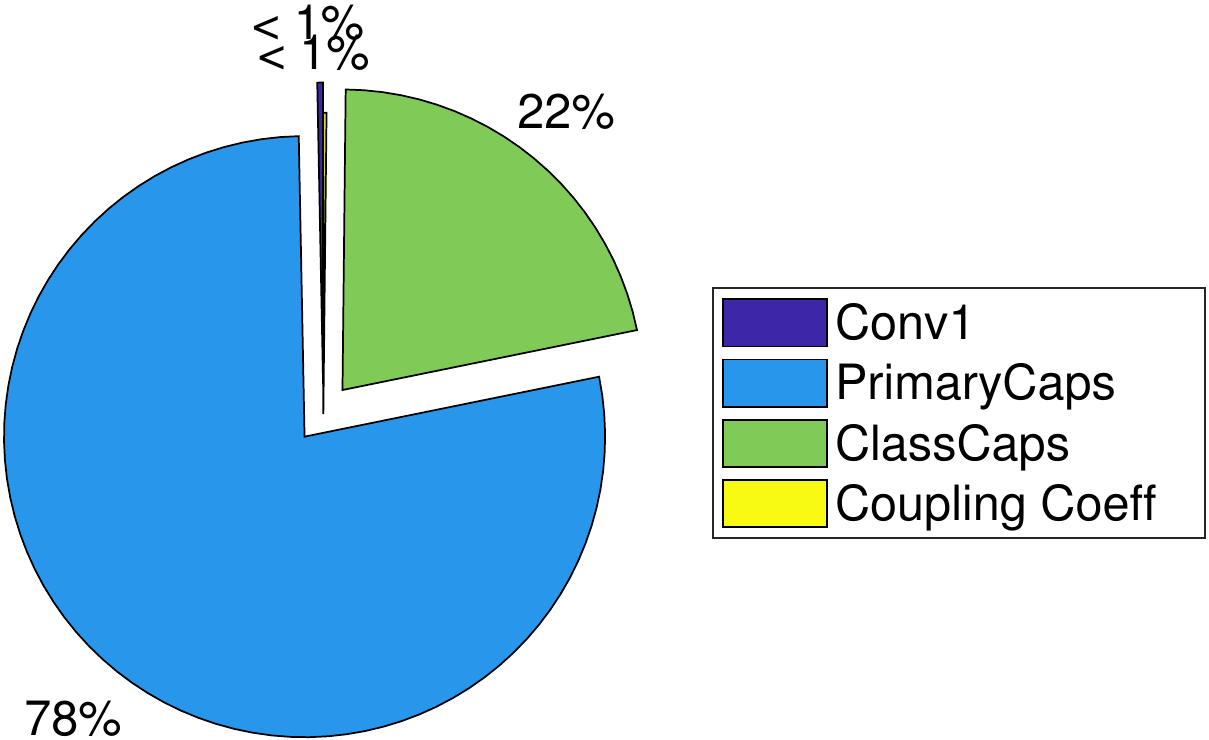}
	\caption{Distribution of trainable parameters on the CapsuleNet across different layers.}
	\label{fig:pie_parameters}
\end{figure}
\end{minipage}
\hspace{.02\linewidth}
\begin{minipage}[t]{.58\linewidth}
\vspace{8mm}
\begin{table}[H]
	\centering
	\resizebox{\linewidth}{!}{%
	\begin{tabular}{c|ccc}
		& \textbf{Inputs} & \textbf{\# parameters} & \textbf{Outputs} \\
		\hline
		Conv1 & 784 & 20992 & 102400 \\
		\hline
		PrimaryCaps & 102400 & 5308672 & 102400 \\
		\hline
		ClassCaps & 102400 & 1474560 & 160 \\
		\hline
		Coupling Coeff & 160 & 11520 & 160
	\end{tabular}
	}
    \vspace{2mm}
	\caption{Input size, number of trainable parameters and output size of each layer of the CapsuleNet.}
	\label{tab:trainable_parameters}
\end{table}
\end{minipage}
\vspace*{0mm}
\end{figure}

\Cref{fig:pie_parameters} shows quantitatively how many parameters are needed for each layer. As evident, the majority of the weights belong to the PrimaryCaps layer, due to its $256$ channels and $8$D capsules. Even if the ClassCaps layer has fully-connected behavior, it counts just for less than $25\%$ of the total parameters of the CapsuleNet. Finally, Conv1 and the coupling coefficients counts for a very small percentage of the parameters. The detailed computation of the parameters is reported in \Cref{tab:trainable_parameters}. Based on that, we make an observation valuable for designing our hardware accelerator: \textit{by considering an $8$-bit fixed point weight representation, we can estimate that an on-chip memory size of $8$MB is large enough to contain every parameter of the CapsuleNet.}

\subsection{Performance Analysis on a GPU}
\label{subsec:performance_GPU}

\begin{figure}[t]
\noindent
\begin{minipage}[t]{.59\linewidth}
\begin{figure}[H]
	\centering
	\includegraphics[width=\linewidth]{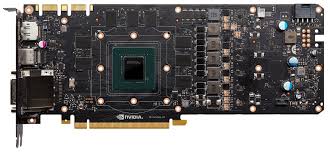}
	\caption{Nvidia Ge-Force GTX1070.}
	\label{fig:GPU_board}
\end{figure}
\end{minipage}
\hfill
\begin{minipage}[t]{.39\linewidth}
\begin{figure}[H]
	\centering
	\includegraphics[width=\linewidth]{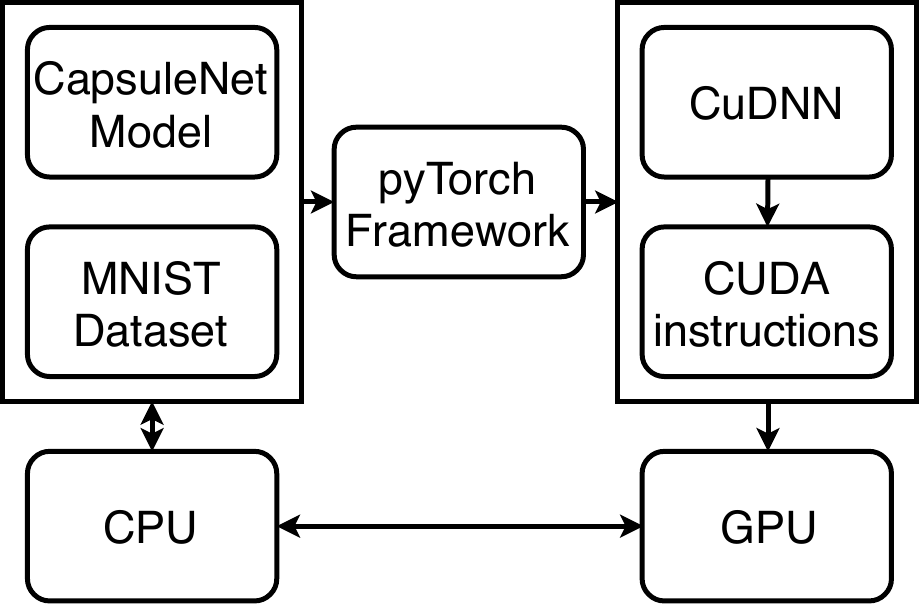}
	\caption{Experimental setup for GPU analyses.}
	\label{fig:GPU_setup}
\end{figure}
\end{minipage}
\vspace*{0mm}
\end{figure}

\begin{figure}[t]
\vspace*{0mm}
\noindent
\begin{minipage}[t]{.49\linewidth}
\begin{figure}[H]
	\centering
	\includegraphics[width=\linewidth]{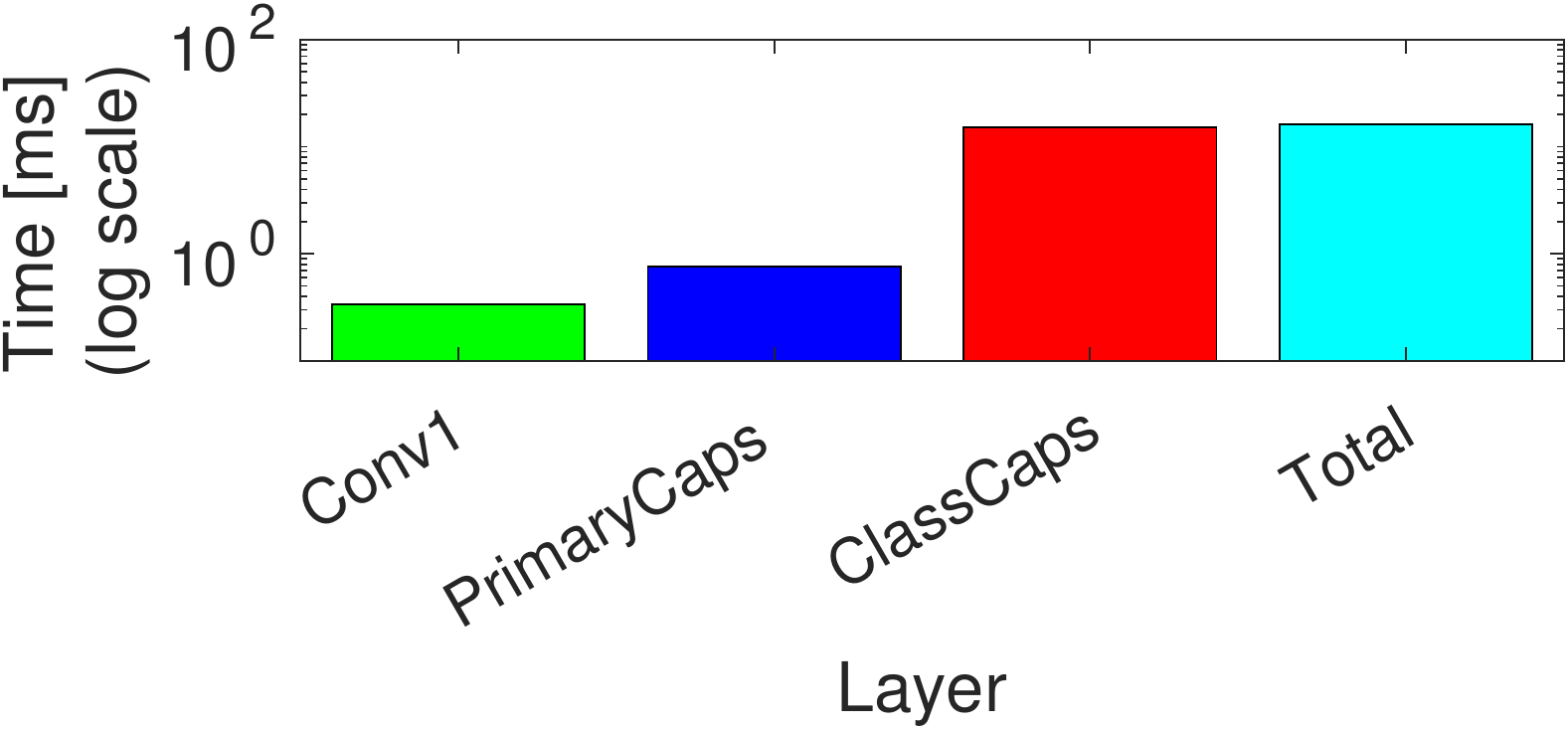}
	\caption{Layer-wise performance of the inference pass of the CapsuleNet.}
	\label{fig:bar_layers}
\end{figure}
\end{minipage}
\hspace{.02\linewidth}
\begin{minipage}[t]{.49\linewidth}
\begin{figure}[H]
	\centering
	\includegraphics[width=\linewidth]{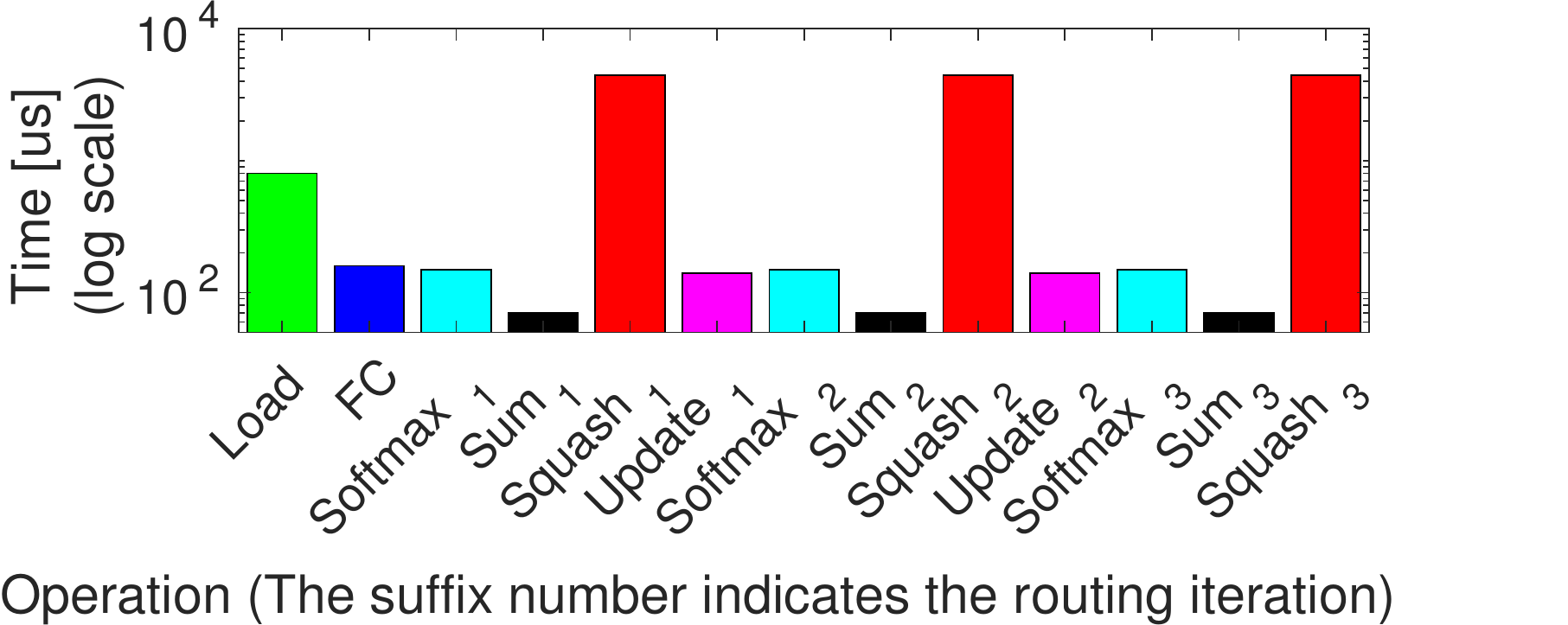}
	\caption{Performance of the inference pass on each step of the routing-by-agreement algorithm.}
	\label{fig:bar_routing}
\end{figure}
\end{minipage}
\vspace*{0mm}
\end{figure}

At this stage, we measure the time required for an inference pass on the GPU. The experimental setup is shown in \Cref{fig:GPU_setup}. \Cref{fig:bar_layers} shows the measurements for each layer. The ClassCaps layer is the computational bottleneck, because it is around $10$$\times$ slower than the previous layers. To obtain more detailed results, a further analysis has been performed, regarding the performance for each step of the routing-by-agreement (\Cref{fig:bar_routing}). It is evident that \textit{the Squashing operation inside the ClassCaps layer represents the most compute-intensive operation}. This analysis gives us the motivation to \textit{spend more effort in optimizing routing-by-agreement and squashing} in our CapsuleNet accelerator.

\subsection{Summary of Key Observations from our Analyses}
\label{subsec:key_observations}

From the analyses performed in \Cref{subsec:trainable_parameters,subsec:performance_GPU}, we derive the following key observations:

\begin{itemize}
\item The CapsuleNet inference performed on GPU is more compute-intensive than memory-intensive, because the bottleneck is represented by the squashing operation.
\item A massive parallel computation capability in the hardware accelerator is desirable to achieve the same or a better level of performance than the GPU for Conv1 and ClassCaps layers.
\item Since the overall memory required to store all the weights is quite high, the buffers located in between the on-chip memory and the processing elements are beneficial to maintain high throughput and to mitigate the latency due to on-chip memory reads.
\end{itemize}

\section{Designing the CapsAcc Architecture}
\label{sec:design}

\begin{figure}[t]
	\centering
	\includegraphics[width=.6\linewidth]{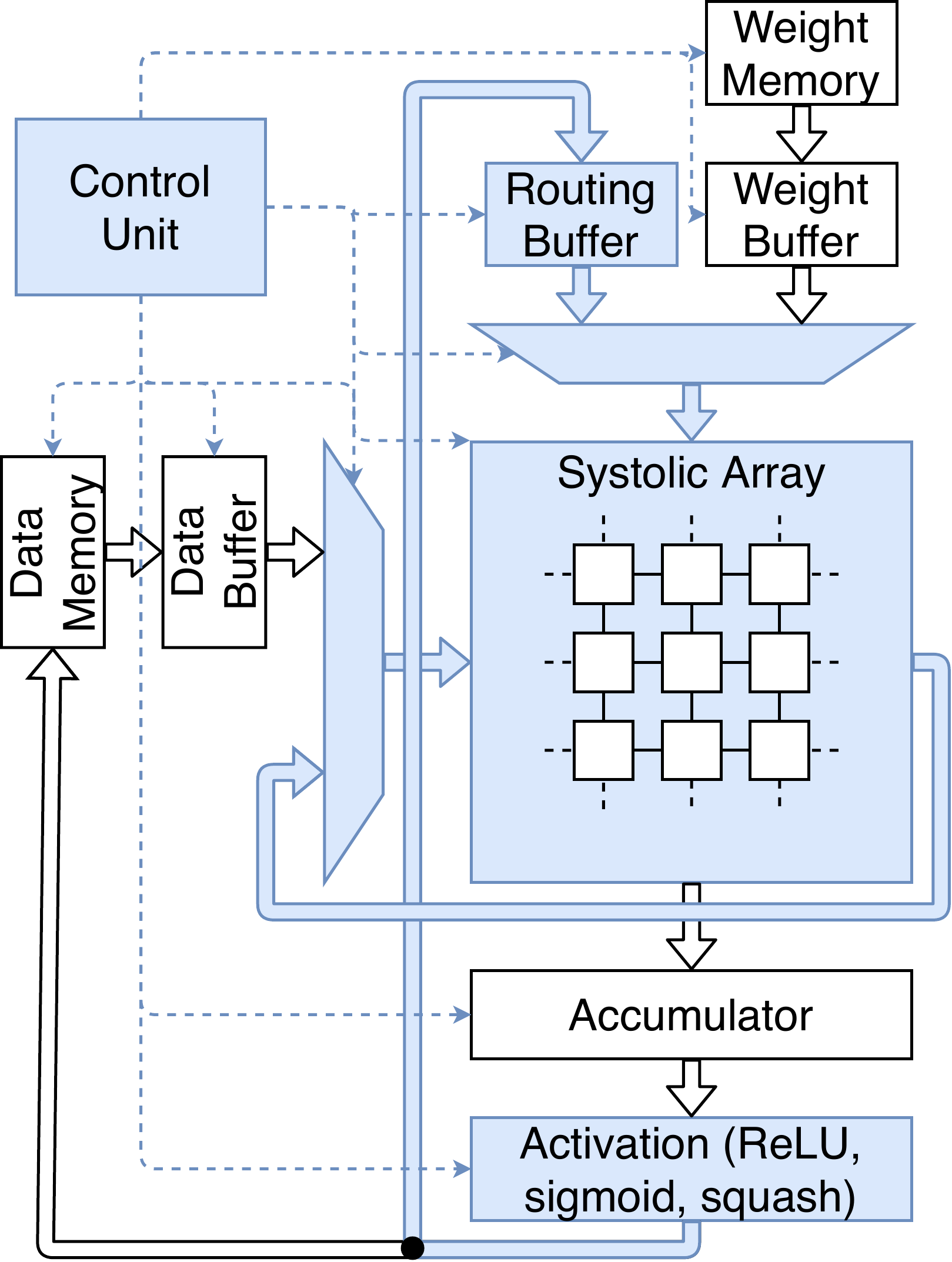}
	\caption{Overview of our CapsAcc Architecture.}
	\label{fig:top_architecture}
    \vspace*{0mm}
\end{figure}

\begin{figure*}[t]
\centering
\begin{minipage}[t]{.24\linewidth}
\vspace*{0mm}
\subfloat[]{
\includegraphics[width=\linewidth]{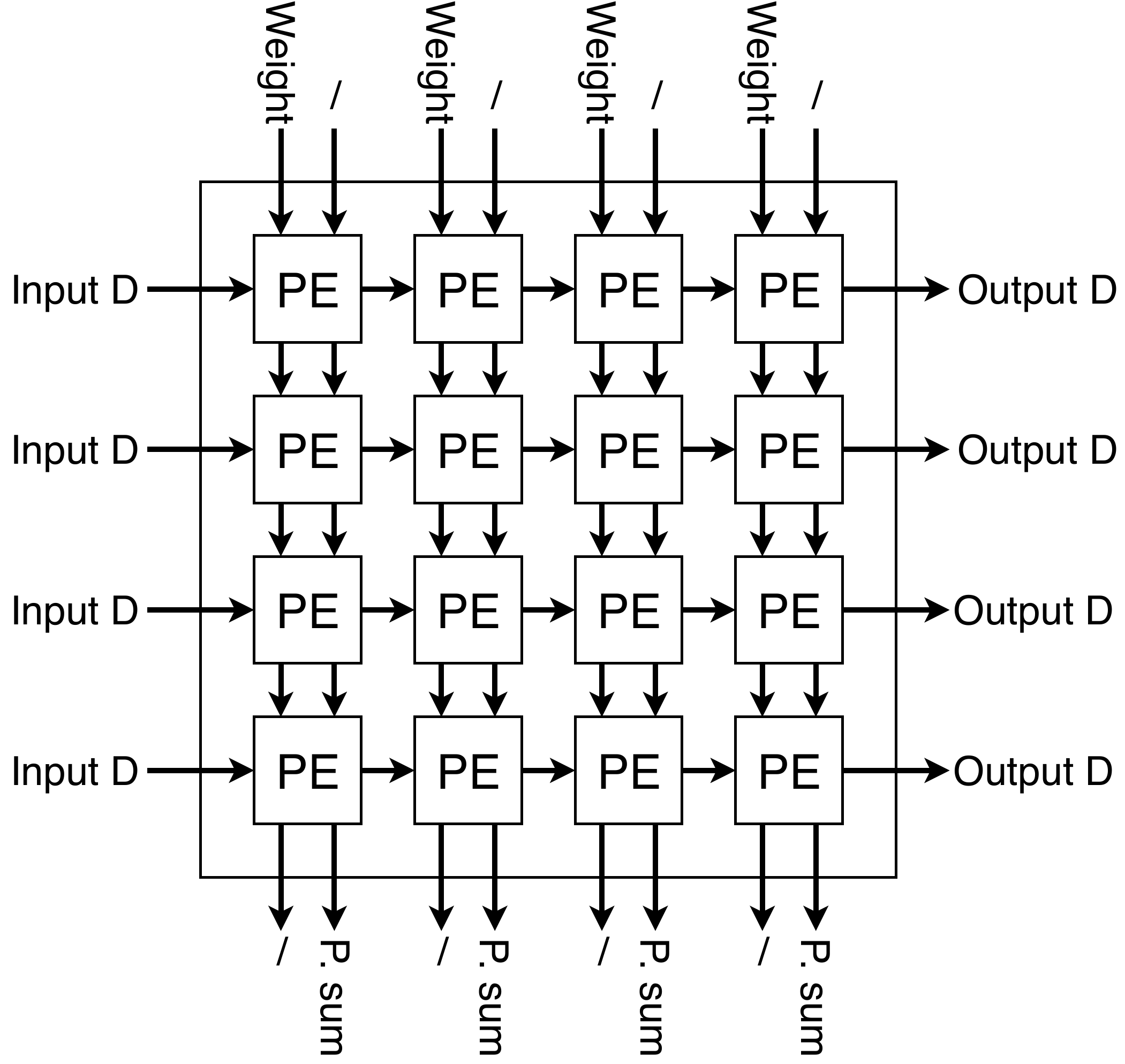}
\label{fig:systolic_array}}
\end{minipage}
\hfill
\begin{minipage}[t]{.18\linewidth}
\vspace*{0mm}
\subfloat[]{
\includegraphics[width=\linewidth]{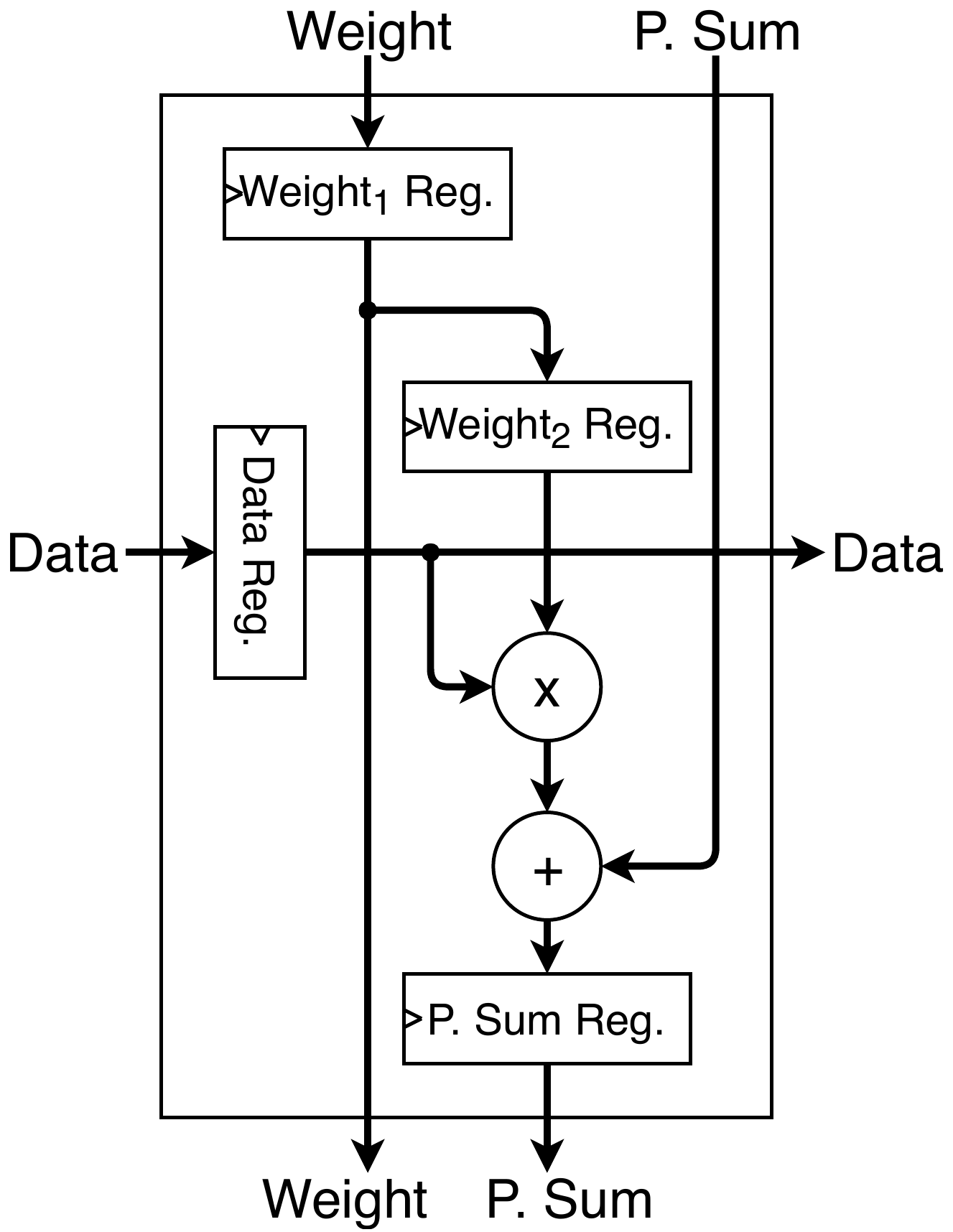}
\label{fig:processing_element}}
\end{minipage}
\hfill
\begin{minipage}[t]{.09\linewidth}
\subfloat[]{
\includegraphics[width=\linewidth]{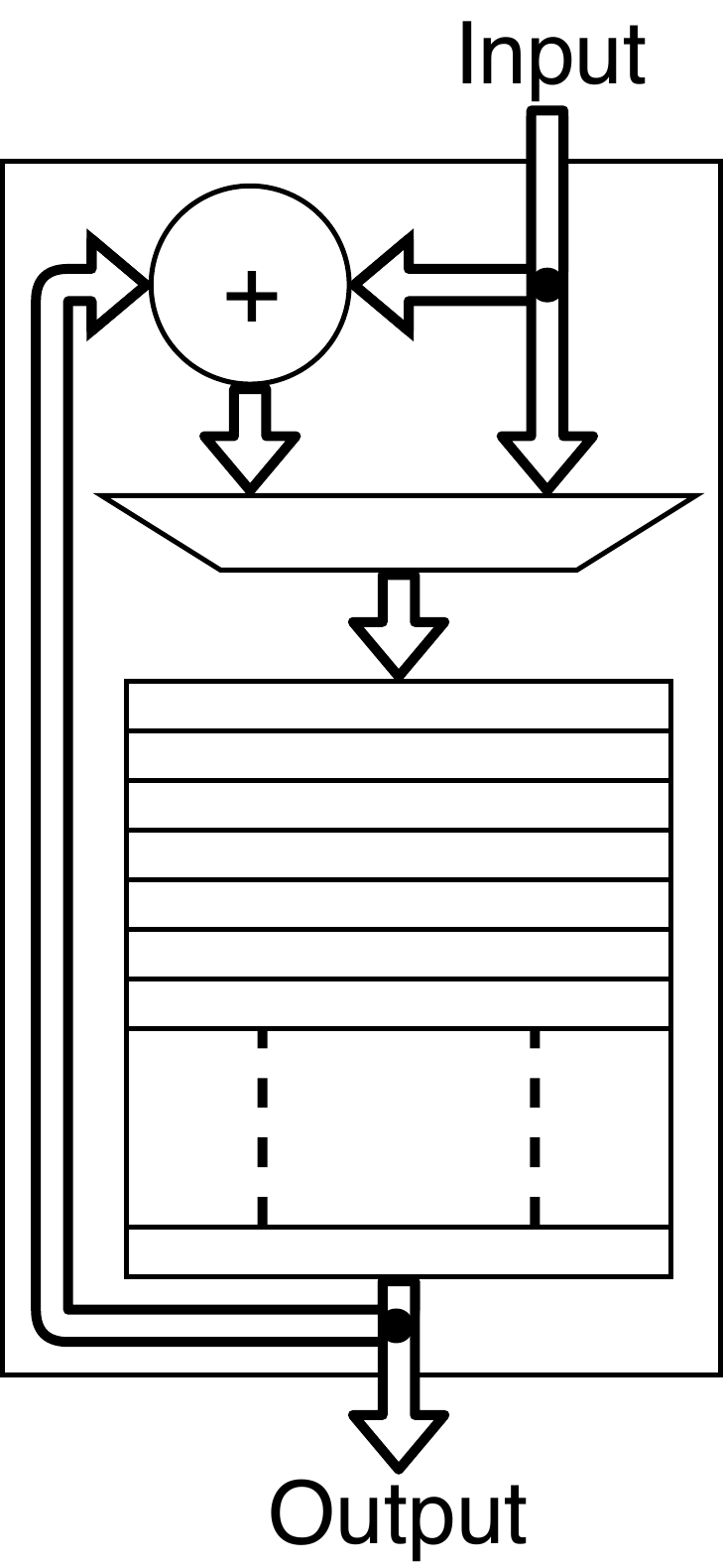}
\label{fig:accumulator}}
\end{minipage}
\hfill
\begin{minipage}[t]{.18\linewidth}
\vspace*{0mm}
\subfloat[]{
\includegraphics[width=\linewidth]{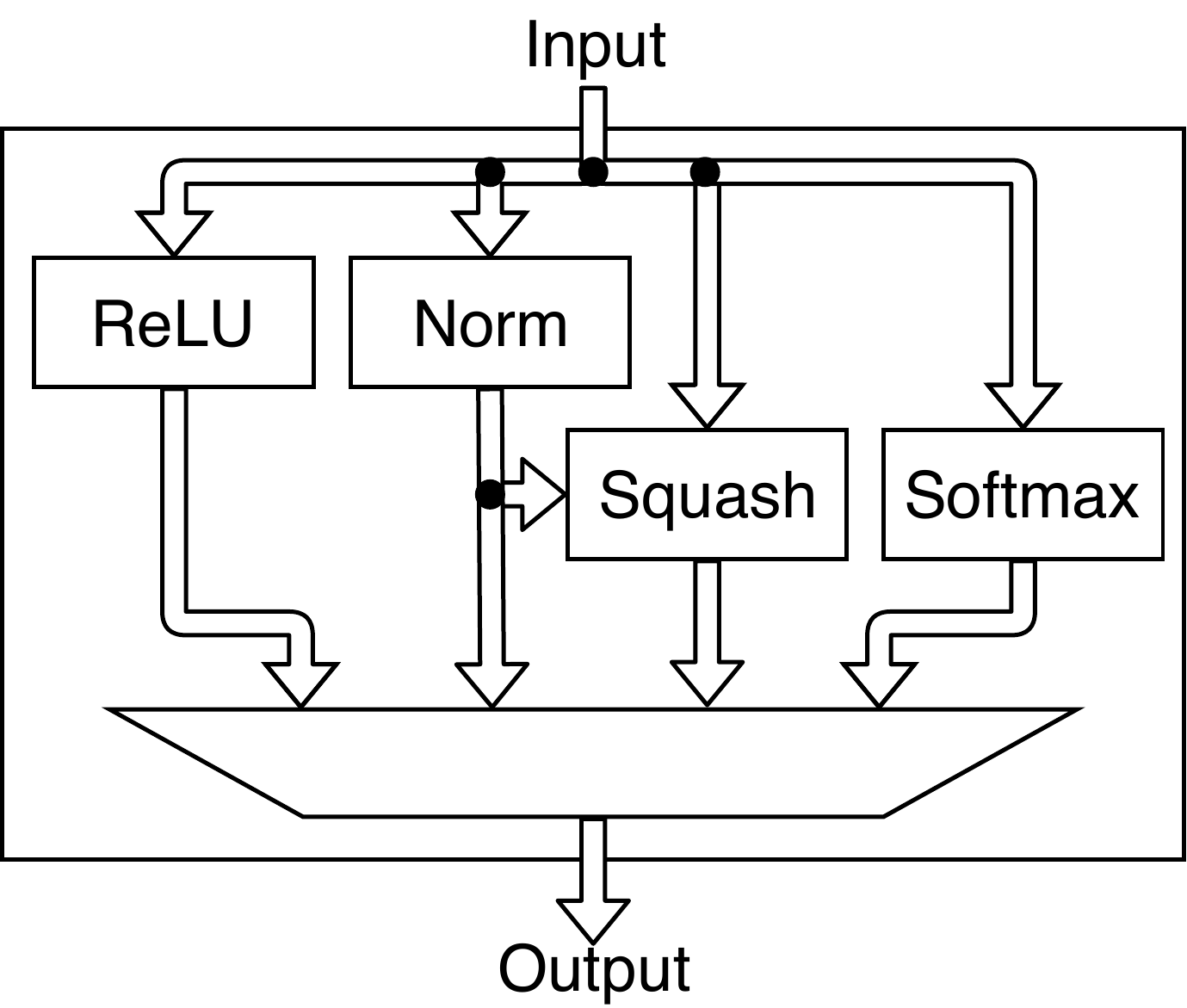}
\label{fig:activation}}
\newline
\vspace*{-4mm}
\centering
\subfloat[]{
\includegraphics[width=.7\linewidth]{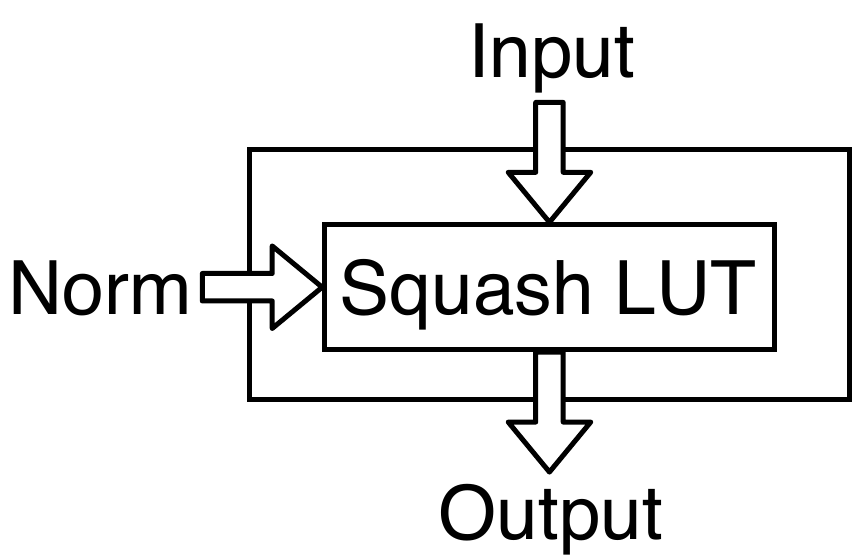}
\label{fig:squash}}
\end{minipage}
\hfill
\begin{minipage}[t]{.10\linewidth}
\subfloat[]{
\includegraphics[width=\linewidth]{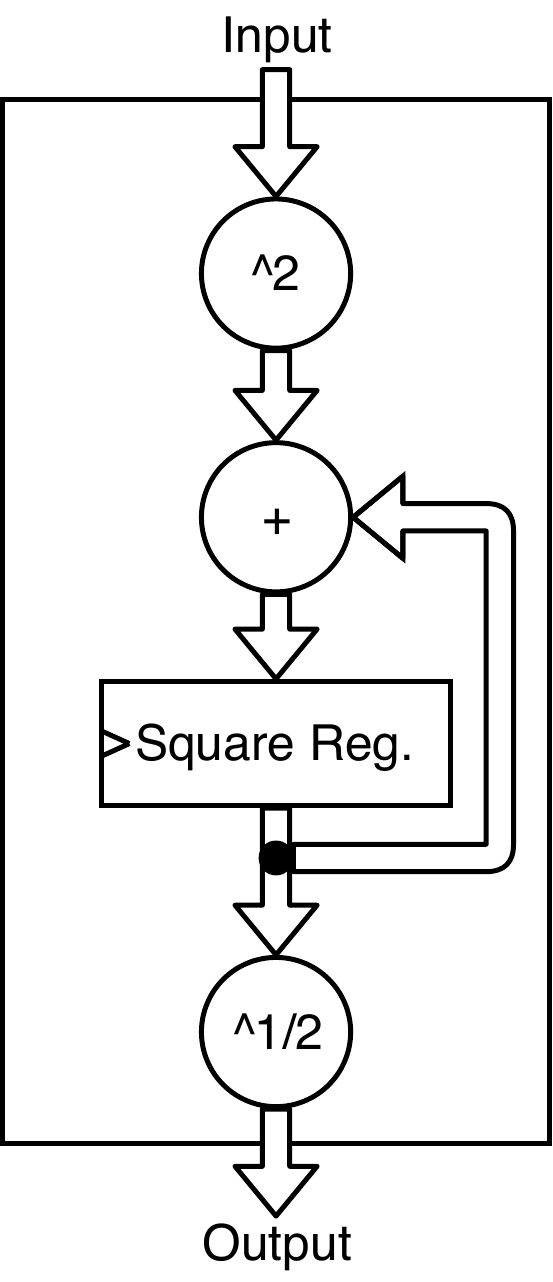}
\label{fig:norm}}
\end{minipage}
\hfill
\begin{minipage}[t]{.10\linewidth}
\subfloat[]{
\includegraphics[width=\linewidth]{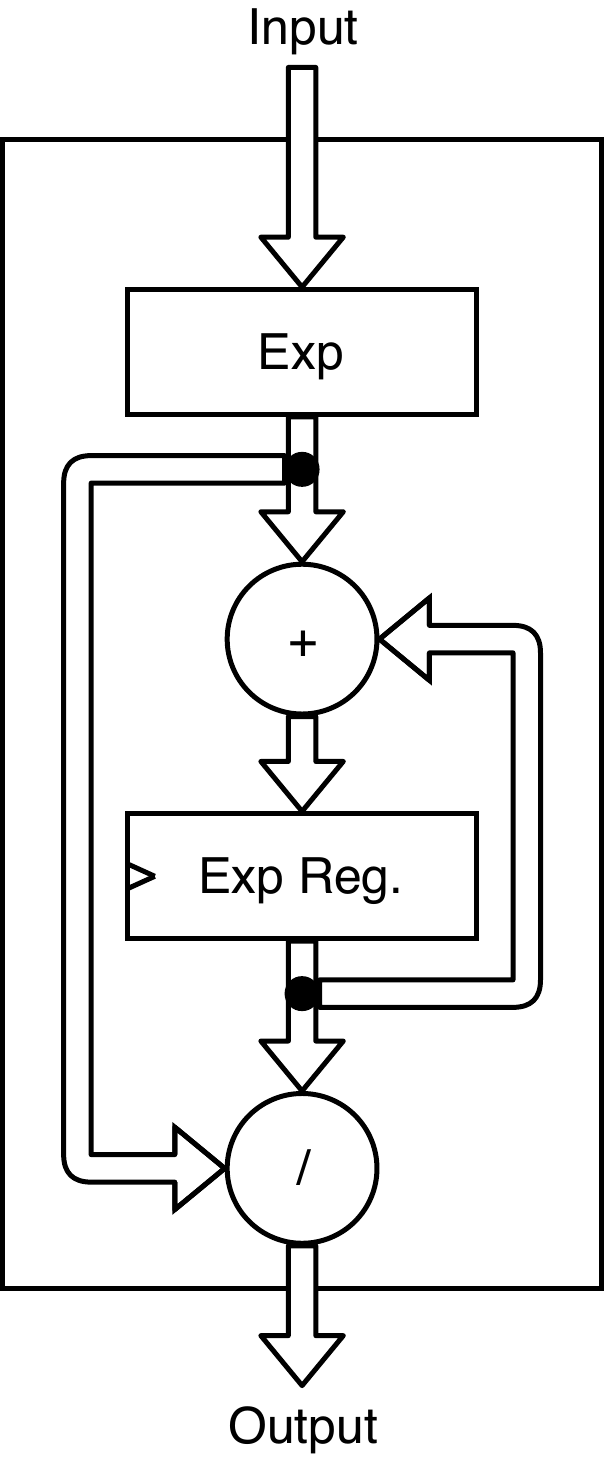}
\label{fig:softmax}}
\end{minipage}
\caption{Architecture of Different Components of our CapsAcc Accelerator: (a) Systolic Array. (b) A Processing Element of the Systolic Array. (c) Accumulator. (d) Activation Unit. (e) Squashing Function Unit. (f) Norm Function Unit. (g) Softmax Function Unit.}
\label{fig:architecture_components}
\vspace*{0mm}
\end{figure*}

Following the above observations, we designed the complete CapsAcc accelerator and implemented it in hardware (RTL). The top-level architecture is shown in \Cref{fig:top_architecture}, where the blue-colored blocks highlight our novel contributions over other existing accelerators for CNNs. The detailed architectures of different components of our accelerator are shown in \Cref{fig:architecture_components}. Our CapsAcc architecture has a systolic array supporting a specialized data-flow mapping (see \Cref{sec:mapping}), which allows to exploit the computational parallelism for multi-dimensional matrix operations. The partial sums are stored and properly added together by the accumulator unit. The activation unit performs different activation functions, according to the requirements for each stage. The buffers (Data, Routing and Weight Buffers) are essential to temporarily store the information to feed the systolic array without accessing every time to the data and weight memories. The two multiplexers in front of the systolic array introduce the flexibility to process new data or reuse them, according to the data-flow mapping. The control unit coordinates all the accelerator operations, at each stage of the inference.

\subsection{Systolic Array}
\label{subsec:systolic_array}

The systolic array of our CapsAcc architecture is shown in \Cref{fig:systolic_array}. It is composed of a $2$D array of Processing Elements (PEs), with $n$ rows and $m$ columns. For illustration and space reasons, \Cref{fig:systolic_array} presents the $4$$\times$$4$ version, while in our actual CapsAcc design we use a $16$$\times$$16$ systolic array. The inputs are propagated towards the outputs of the systolic array both horizontally (Data) and vertically (Weight, Partial sum). In the first row, the inputs corresponding to the Partial sums are zero-valued, because each sum at this stage is equal to $0$. Meanwhile, the Weight outputs in the last row are not connected, because they  are not used in the following stages.

\Cref{fig:processing_element} shows the data path of a single Processing Element (PE). It has $3$ inputs and $3$ outputs: Data, Weight and Partial sum, respectively. The core of the PE is composed of the sequence of a multiplier and an adder. As shown in \Cref{fig:processing_element}, it has $4$ internal registers: 
(1) {\em Data Reg.} to store and synchronize the Data value coming from the left; 
(2) {\em Sum Reg.} to store the Partial sum before sending it to the neighbor PE below; 
(3) {\em Weight$_1$ Reg.} synchronizes the vertical transfer; 
(4) {\em Weight$_2$ Reg.} stores the value for data reuse. 
The latter is particularly useful for convolutional layers, where the same weight of the filter must be convolved across different data. For fully-connected computations, the second weight register introduces just one clock cycle latency, without affecting the throughput. The bit-widths of each element have been designed as follows: 
(1) each PE computes the product between an $8$-bit fixed-point Data and an $8$-bit fixed-point Weight; and 
(2) the sum is designed as a $25$-bit fixed-point value. 
At full throttle, each PE produces one output-per-clock cycle, which also implies one output-per-clock cycle for every column of the systolic array.

\subsection{Accumulator}
\label{subsec:accumulator}

The Accumulator unit consists of a FIFO buffer to store the Partial sums coming from the systolic array, and sum them together when needed. The multiplexer allows the choice to feed the buffer with the data coming from the systolic array or with the one coming from the internal adder of the Accumulator. We designed the Accumulator to have $25$-bit fixed-point data. \Cref{fig:accumulator} shows the data path of our Accumulator. In the overall CapsAcc there are as many Accumulators as the number of columns of the systolic array.

\subsection{Activation Unit}
\label{subsec:activation}

The Activation Unit follows the Accumulators. As shown in \Cref{fig:activation}, it performs different functions in parallel, while the multiplexer (placed at the bottom of the figure) selects the path to propagate the information towards the output. As for the case of the Accumulator, the figure shows only one unit, while in the complete CapsAcc architecture there is one Activation Unit per each column of the systolic array. The $25$-bits data values coming from the Accumulators are reduced to an $8$-bit fixed-point value, to reduce the computations at this stage.

Note: the Rectified Linear Unit (ReLU) \cite{ref:ReLU} is a very simple function and its implementation description is omitted, since it is straightforward. This function is used for every feature of the first two layers of the CapsuleNet.

We designed the \textbf{Normalization operator (Norm)} with a structure similar to the Multiply-and-Accumulate operator, where, instead of a traditional multiplier, there is the {\em Power2} operator. Its data path is shown in \Cref{fig:norm}. A register stores the partial sum and the {\em Sqrt} operator produces the output. We designed the square operator as a Look Up Table with $12$-bit input and $8$-bit output. It produces a valid output every $n+1$ clock cycles, where $n$ is the size of the array for which we want to compute the Norm. This operator is used either as it is to compute the classification prediction, or as an input for the Squashing function.

We designed and implemented the \textbf{Squashing function} as a Look Up Table, as shown in \Cref{fig:squash}. Looking at \Cref{eq:squash}, the function takes an input $s_j$ and its norm $\left| \left| s_j \right| \right|$. The Norm input is coming from its respective unit. Hence, this Norm operation is not implemented again inside the Squash unit. The LUT takes as input a $6$-bit fixed-point data and a $5$-bit fixed-point norm to produce an $8$-bit output. We decided to limit the bit-width to reduce the computational requirements at this stage, following the analysis performed in \Cref{sec:analysis} that shows the highest computational load for this operation. A valid output is produced with just one additional clock cycle compared to the Norm.

The \textbf{Softmax function} design is shown in \Cref{fig:softmax}. First, it computes the exponential function ($8$-bit Look Up Table) and accumulates the sum in a register, followed by division. Overall, having an array of $n$ elements, this block is able to compute the softmax function of the whole array in $2n$ clock cycles.

\subsection{Control Unit}
\label{subsec:control_unit}

At each stage of the inference process, it generates different control signals for all the components of the accelerator architecture, according to the operations needed. It is essential for correct operation of the accelerator.

\section{Data-Flow Mapping}
\label{sec:mapping}

\begin{figure*}[t]
\begin{minipage}[t]{.24\linewidth}
\subfloat[]{
\includegraphics[width=\linewidth]{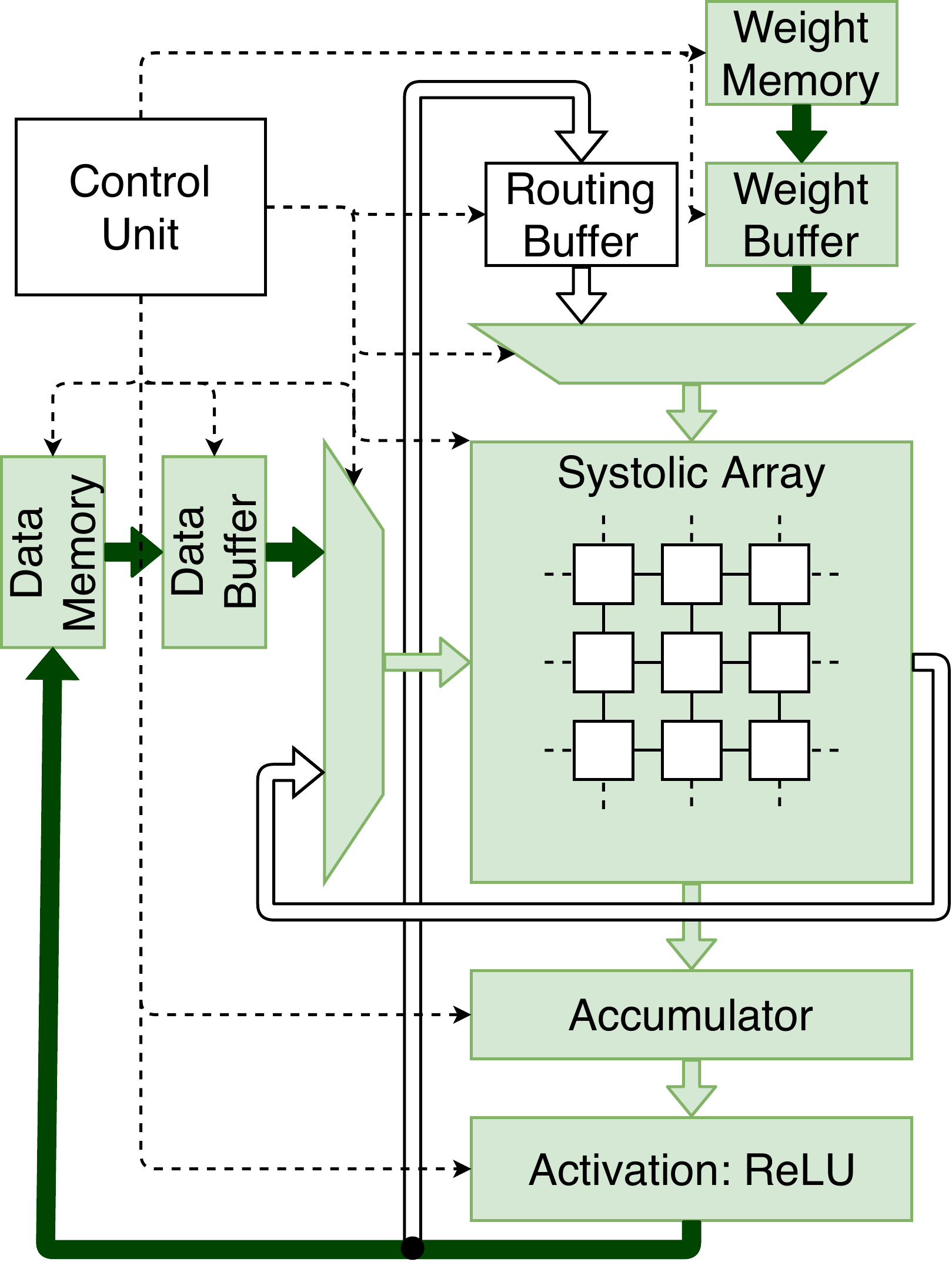}
\label{fig:top_conv}}
\end{minipage}
\hfill
\begin{minipage}[t]{.24\linewidth}
\subfloat[]{
\includegraphics[width=\linewidth]{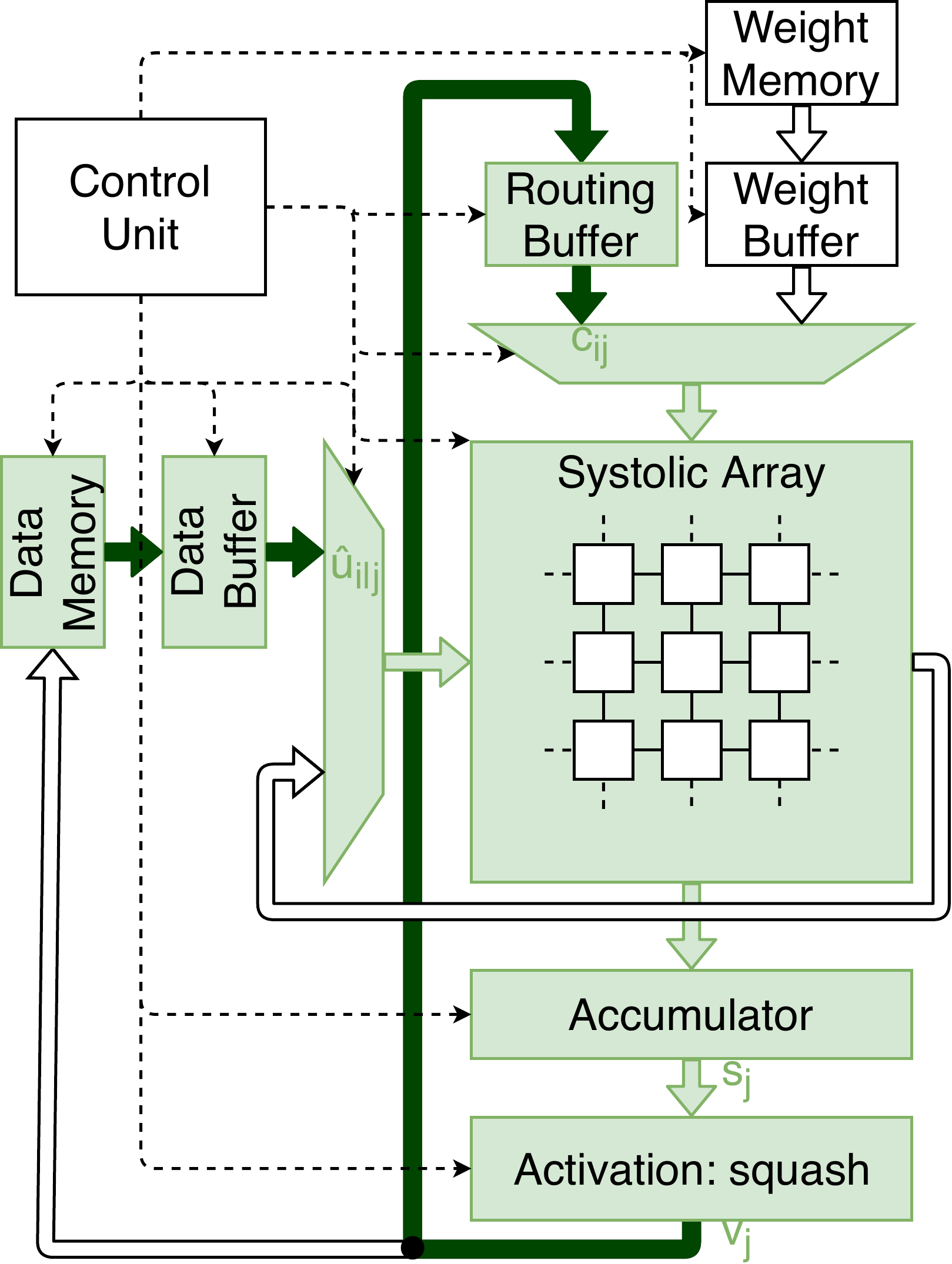}
\label{fig:top_routing_squash_first}}
\end{minipage}
\hfill
\begin{minipage}[t]{.24\linewidth}
\subfloat[]{
\includegraphics[width=\linewidth]{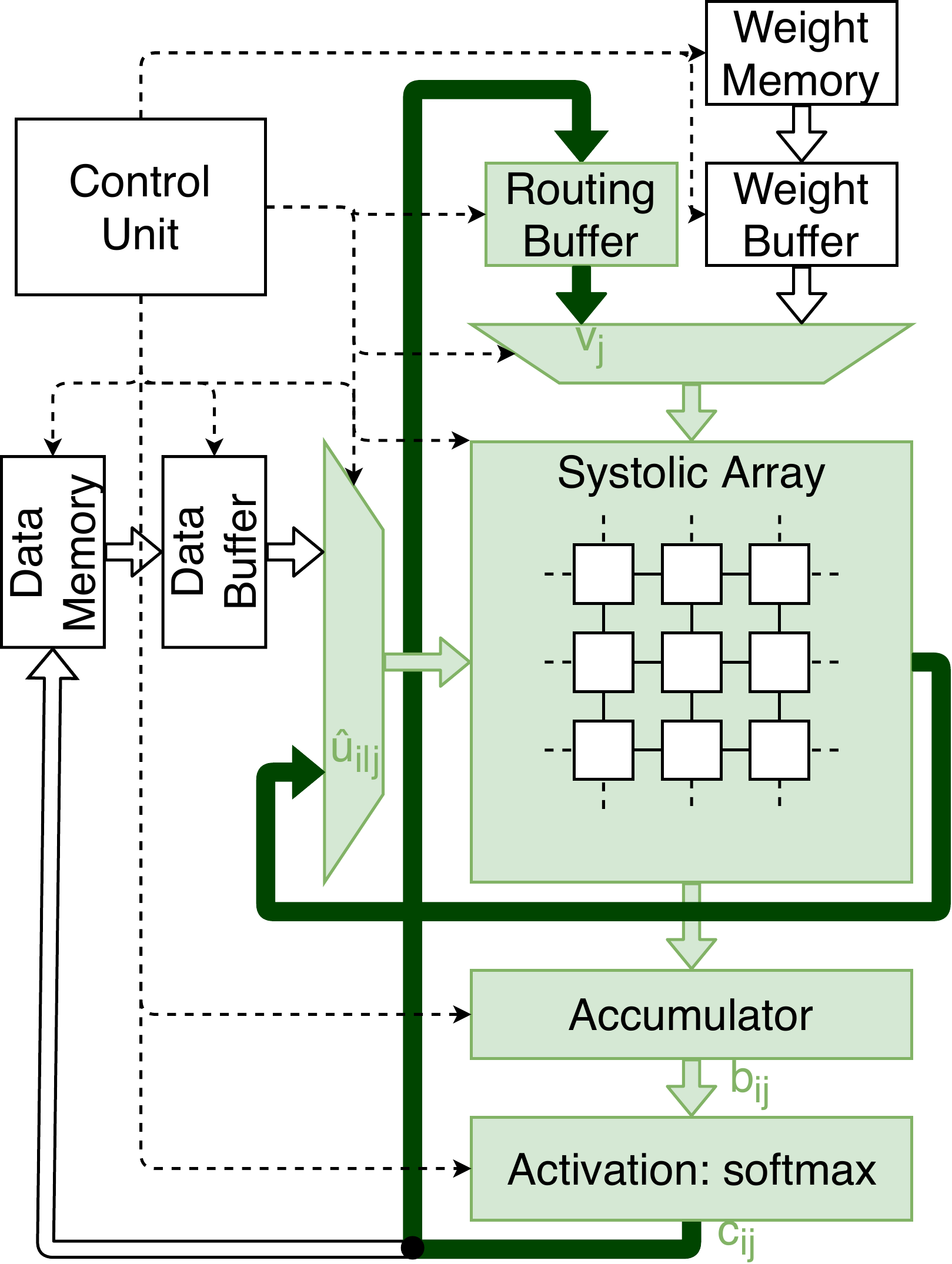}
\label{fig:top_routing_softmax}}
\end{minipage}
\hfill
\begin{minipage}[t]{.24\linewidth}
\subfloat[]{
\includegraphics[width=\linewidth]{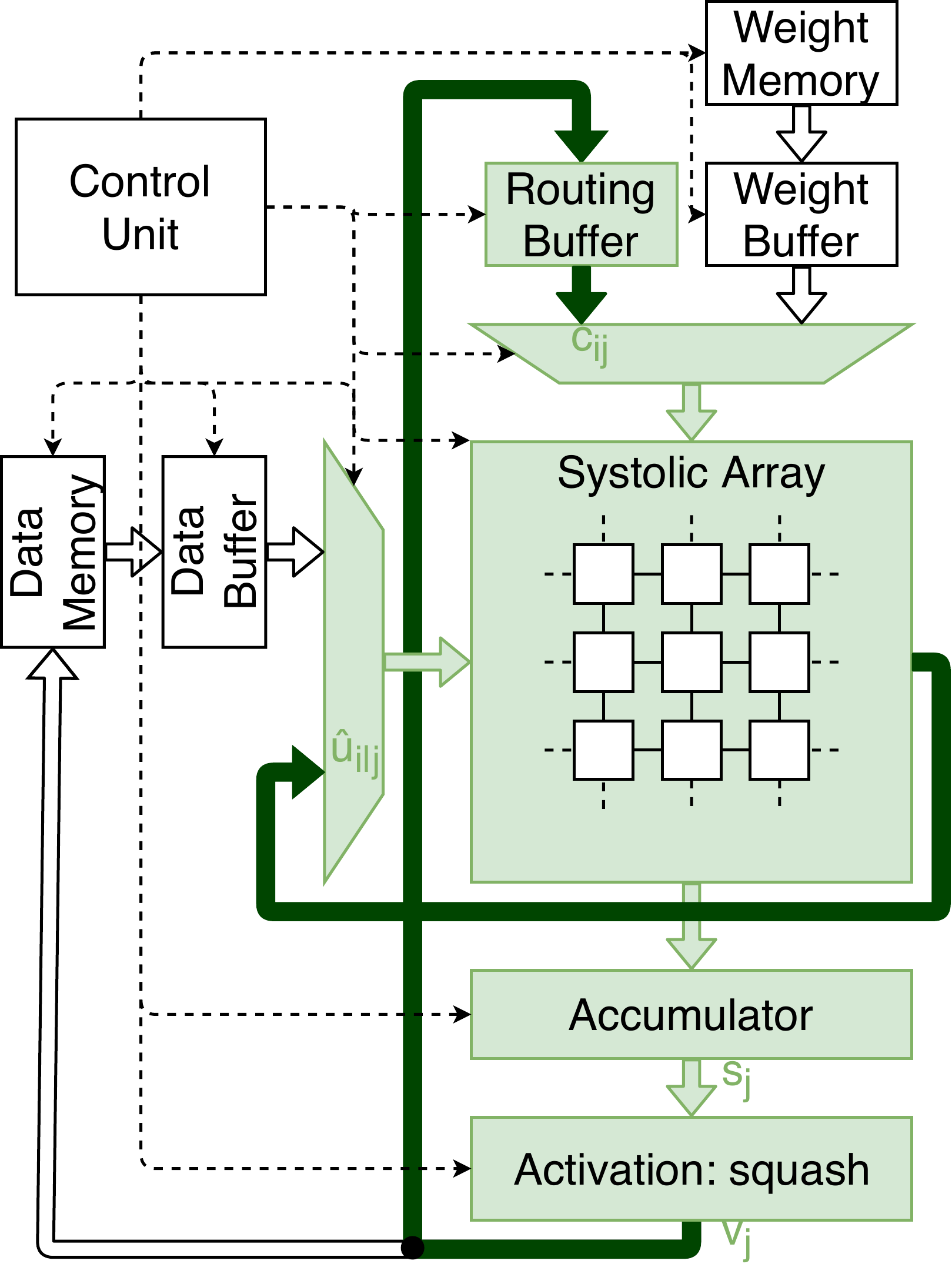}
\label{fig:top_routing_squash}}
\end{minipage}
\label{fig:mapping_example}
\caption{Data-flow mapping onto our CapsAcc accelerator for different scenarios of the case study. (a) Convolutional layer mapping. (b) First sum generation \& squashing operation mapping. (c) Update and softmax operation mapping. (d) The sum generation \& squashing operation mapping other than the first routing iteration.}
\vspace*{0mm}
\end{figure*}

\begin{figure}[t]
\begin{algorithmic}[1]
{\small
\STATE \textcolor{blue}{for}(l=0; l$<$L; l++) \textcolor{darkgreen}{//output capsules}
\STATE \ \textcolor{blue}{for}(k=0; k$<$K; k++) \textcolor{darkgreen}{//output channels}
\STATE \ \ \textcolor{blue}{for}(j=0; j$<$J; j++) \textcolor{darkgreen}{//input capsules}
\STATE \ \ \ \textcolor{blue}{for}(i=0; i$<$I; i++) \textcolor{darkgreen}{//input channels}
\STATE \ \ \ \ \textcolor{blue}{for}(g=0; g$<$G; G++) \textcolor{darkgreen}{//output columns in a feature map}
\STATE \ \ \ \ \ \textcolor{blue}{for}(f=0; f$<$F; f++) \textcolor{darkgreen}{//output rows in a feature map}
\STATE \ \ \ \ \ \ \textcolor{blue}{for}(c=0; c$<$C; c++) \textcolor{darkgreen}{//kernel/input columns}
\STATE \ \ \ \ \ \ \ \textcolor{blue}{for}(r=0; r$<$R; r++) \textcolor{darkgreen}{//kernel/input rows}
\STATE \ \ \ \ \ \ \ \ Sum += Weight$\cdot$Data \textcolor{darkgreen}{//multiply and accumulate}
}
\end{algorithmic}
\caption{Mapping algorithm for CapsuleNet operations onto the systolic array.}
\label{alg:mapping}
\vspace*{0mm}
\end{figure}

\begin{figure}[t]
\noindent
\begin{minipage}[b]{.3\linewidth}
\subfloat[]{
\includegraphics[width=\linewidth]{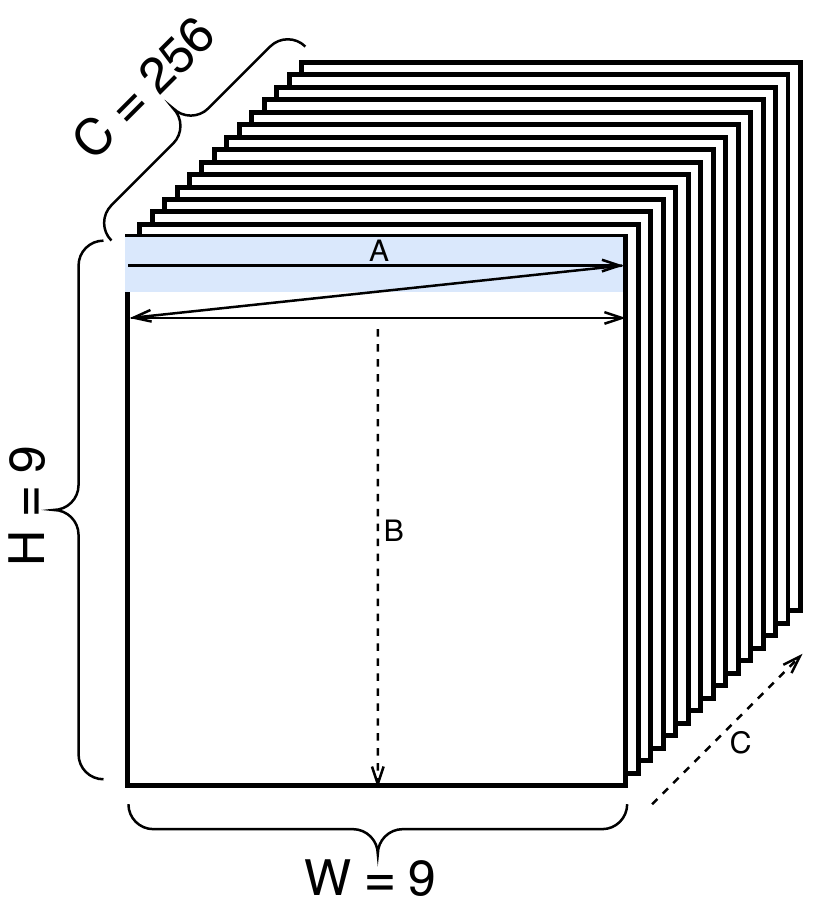}
\label{fig:mapping_conv1}}
\end{minipage}
\hfill
\begin{minipage}[b]{.33\linewidth}
\subfloat[]{
\includegraphics[width=\linewidth]{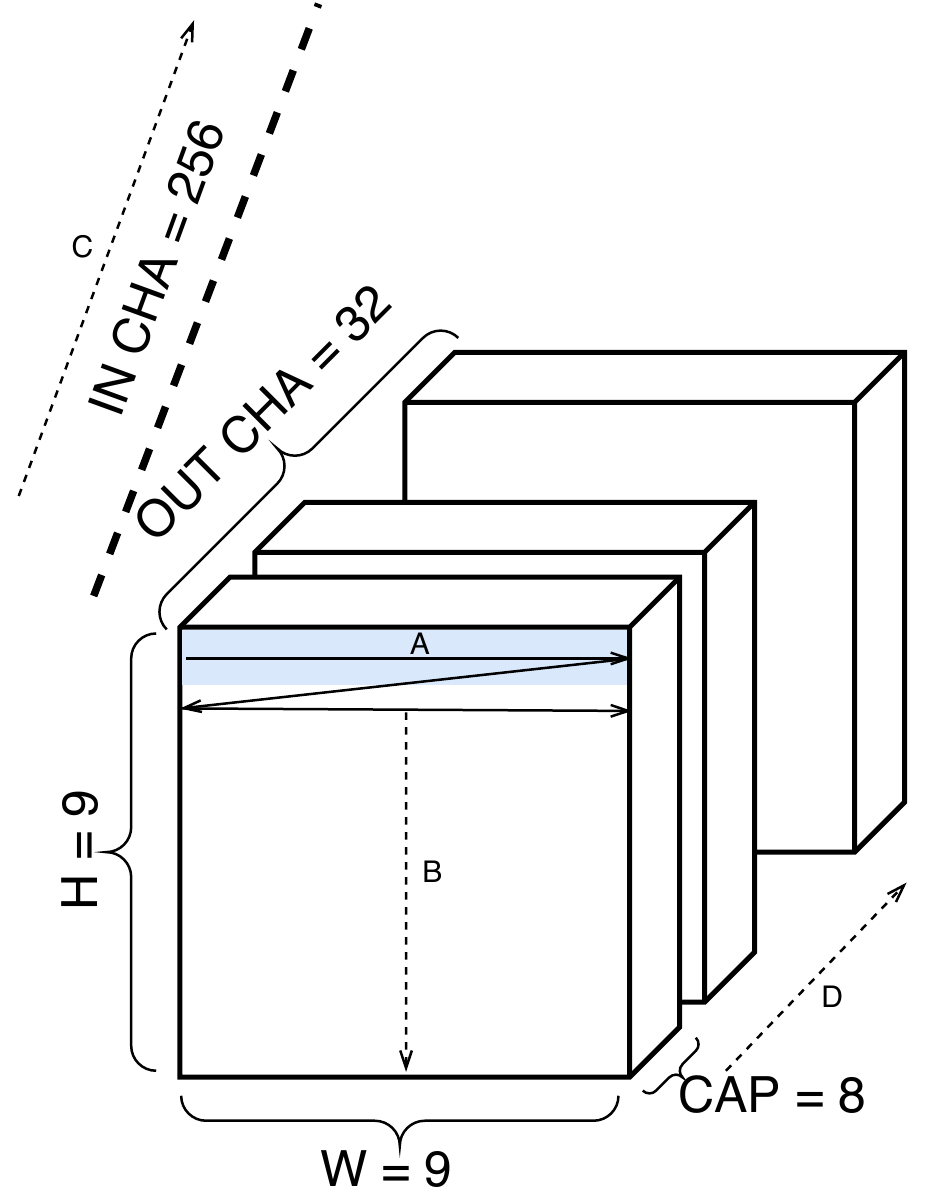}
\label{fig:mapping_primary}}
\end{minipage}
\hfill
\begin{minipage}[b]{.33\linewidth}
\subfloat[]{
\includegraphics[width=\linewidth]{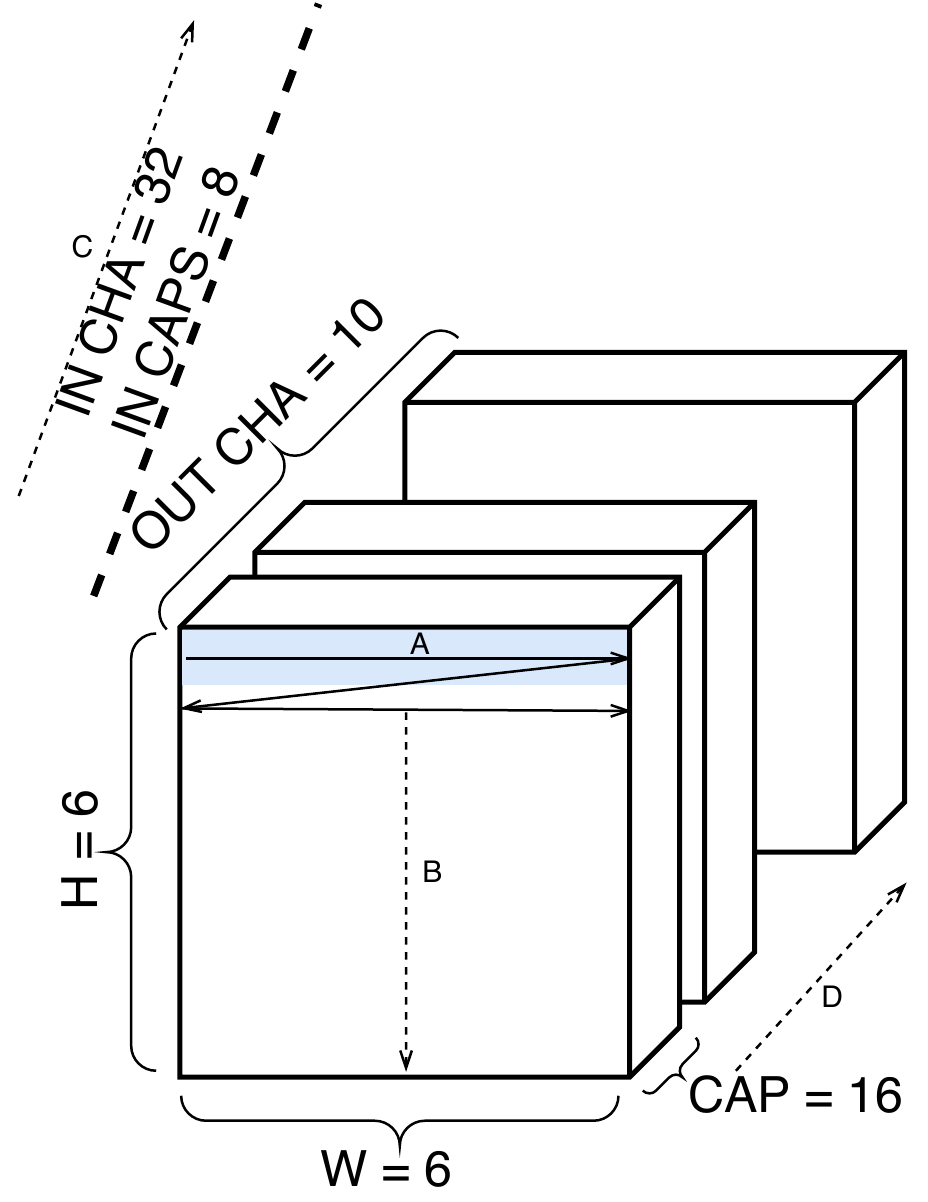}
\label{fig:mapping_digit}}
\end{minipage}
\label{fig:mapping_layers}
\caption{Overview of the process of mapping different layers onto our CapsAcc accelerator. (a) Conv1 layer. (b) PrimaryCaps layer. (c) ClassCaps layer.}
\vspace*{0mm}
\end{figure}

In this section, we provide the details on how to map the processing of different types of layers and operations onto our CapsAcc accelerator, in a step-by-step fashion. To feed the systolic array, we adopt the mapping policy described in \Cref{alg:mapping}. For the ease of understanding, we illustrate the process with the help of an example performing MNIST classification on our CapsAcc accelerator, which also represents our case study. Note, each stage of the CapsuleNet inference requires its own mapping scheme.

\subsection{Conv1 mapping}
\label{subsec:conv1_mapping}

The Conv1 layer has filters of size $9$$\times$$9$ and $256$ channels. As shown in \Cref{fig:mapping_conv1}, we designed the mapping row by row (A,B), and after the last row we move to the next channel (C). \Cref{fig:top_conv} shows how the data-flow is mapped onto our CapsAcc accelerator. To perform the convolution efficiently, we hold the weight values into the systolic array to reuse the filter across different input data.

\subsection{PrimaryCaps mapping}
\label{subsec:primary_mapping}

Compared to the Conv1 layer, the PrimaryCaps layer has one more dimension, which is the capsule size (i.e., $8$). However, we treat the $8$D capsule as a convolutional layer with $8$ output channels. Thus, \Cref{fig:mapping_primary} shows that we map the parameters row-by-row (A,B), then moving through different input channels (C), and only at the third stage we move on to the next output channel (D). This mapping procedure allows us to minimize the accumulator size, because our CapsAcc accelerator computes first the output features for the same output channel. Since the type of this layer is convolutional, the data-flow is the same as the one in the previous layer, as reported in \Cref{fig:top_conv}.

\subsection{ClassCaps mapping}
\label{subsec:class_mapping}

The mapping of the ClassCaps layer is shown in \Cref{fig:mapping_digit}. After mapping row by row (A,B), we consider input capsules and input channels as the third dimension (C), and output capsules and output channels as the fourth dimension (D).

Then, for each step of the routing-by-agreement process, we design the corresponding data-flow mapping. It is a critical phase, because a less efficient mapping can potentially have a huge impact on the overall performance.

First, \textit{we apply an algorithmic optimization on the routing-by-agreement algorithm.} During the first operation, instead of initializing $b_{ij}$ to $0$ and computing the {\em softmax} on them, we directly initialize the coupling coefficients $c_{ij}$. The starting point is indicated with the blue arrow in \Cref{fig:routing}. With this optimization, we can skip the {\em softmax} computation at the first routing iteration. In fact, this operation is dummy, because all the inputs are equal to $0$, thus they do not depend on the current data.

Regarding the data-flow mapping in our CapsAcc accelerator, we can identify three different data-flow scenarios during the routing-by-agreement algorithm:
\begin{enumerate}[label=\textbf{\arabic*})]
	\item \textbf{First sum generation and squash:} The predictions $\hat{u}_{j|i}$ are loaded from the Data Buffer, the coupling coefficients $c_{ij}$ are coming from the Routing Buffer, the systolic array computes the sums $s_j$, the Activation Unit computes and selects Squash, and the outputs $v_j$ are stored back in the Routing Buffer. This data-flow is shown in \Cref{fig:top_routing_squash_first}.

	\item \textbf{Update and softmax:} The predictions $\hat{u}_{j|i}$ are reused through the horizontal feedback of the architecture, $v_j$ are coming from the Routing Buffer, the systolic array computes the updates for $b_{ij}$, and the Softmax at the Activation Unit produces $c_{ij}$ that are stored back in the Routing Buffer. \Cref{fig:top_routing_softmax} shows the data-flow described above.
    
	\item \textbf{Sum generation and Squash:} \Cref{fig:top_routing_squash} shows the data-flow mapping for this scenario. Compared to the \Cref{fig:top_routing_squash_first}, the predictions $\hat{u}_{j|i}$ are coming from the horizontal feedback link, thus exploiting data reuse also in this stage.
    
\end{enumerate}

\section{Results and Discussion}
\label{sec:evaluation}

\subsection{Experimental Setup}
\label{subsec:experimental_setup}

\begin{figure}[t]
	\centering
	\includegraphics[width=.8\linewidth]{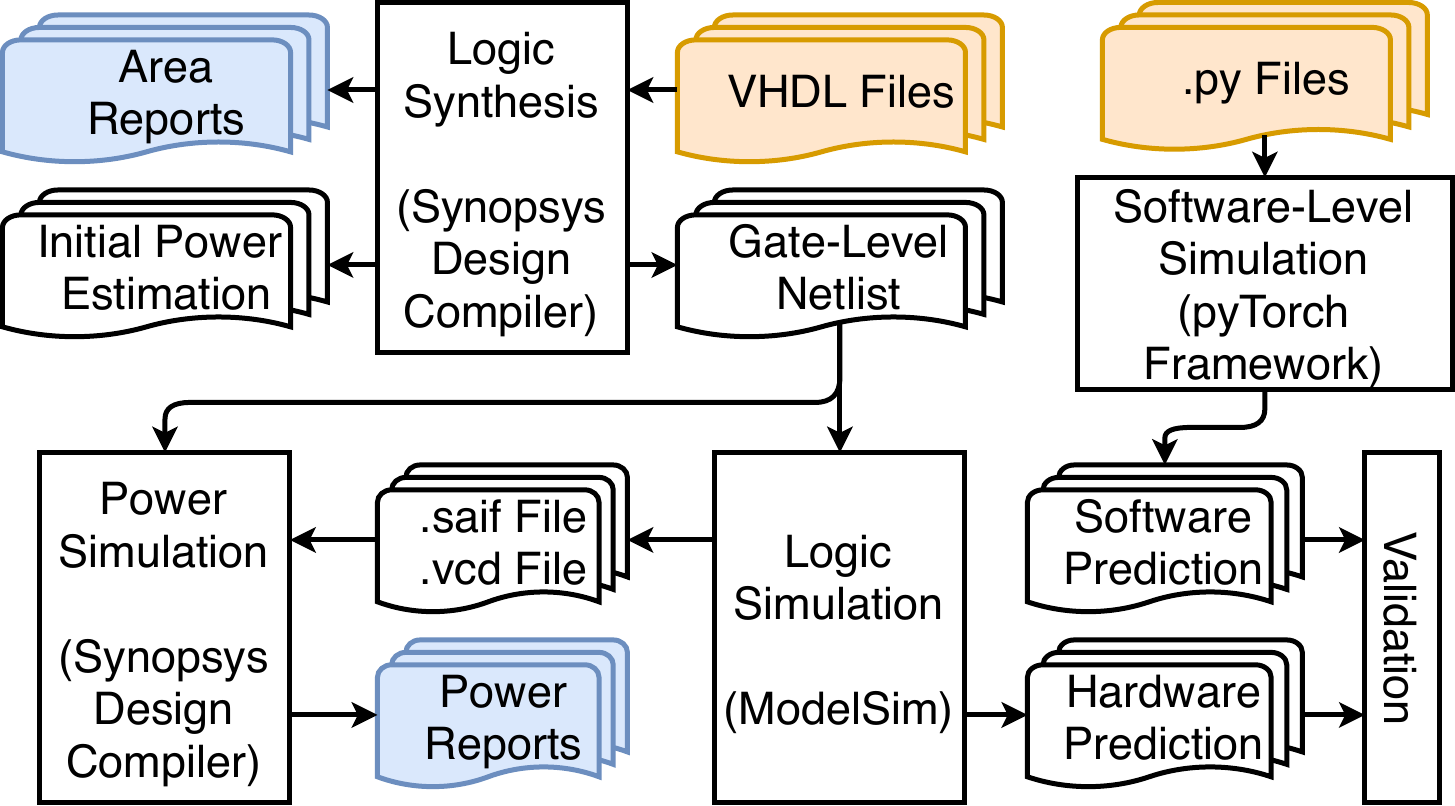}
	\caption{Synthesis flow and tool chain of our experimental setup.}
	\label{fig:synthesis}
\vspace*{0mm}
\end{figure}

We implemented the complete design of our CapsAcc architecture in RTL (VHDL), and evaluated it for the MNIST dataset {\em (to stay consistent with the original CapsuleNet paper)}. We synthesized the complete architecture in a $32$nm CMOS technology library using the ASIC design flow with the Synopsys Design Compiler. We did functional and timing validation through the gate-level simulations using ModelSim, and obtained the precise area, power and performance of our design. The complete synthesis flow is shown in \Cref{fig:synthesis}, where the orange and blue colored boxes represent the inputs and the output results of our experiments, respectively.

{\bf Important Note:} since our hardware design is fully functionally compliant with the original CapsuleNet design of the work of \cite{ref:dyn_routing}, we observed the same accuracy of classification. Therefore, we do not present any classification results in this paper, and only focus on the performance, area and power results, which are more relevant for an optimized hardware architecture.

\subsection{Discussion on Comparative Results}
\label{subsec_discussion}

\begin{figure}[t]
\vspace*{4mm}
	\centering
	\includegraphics[width=.8\linewidth]{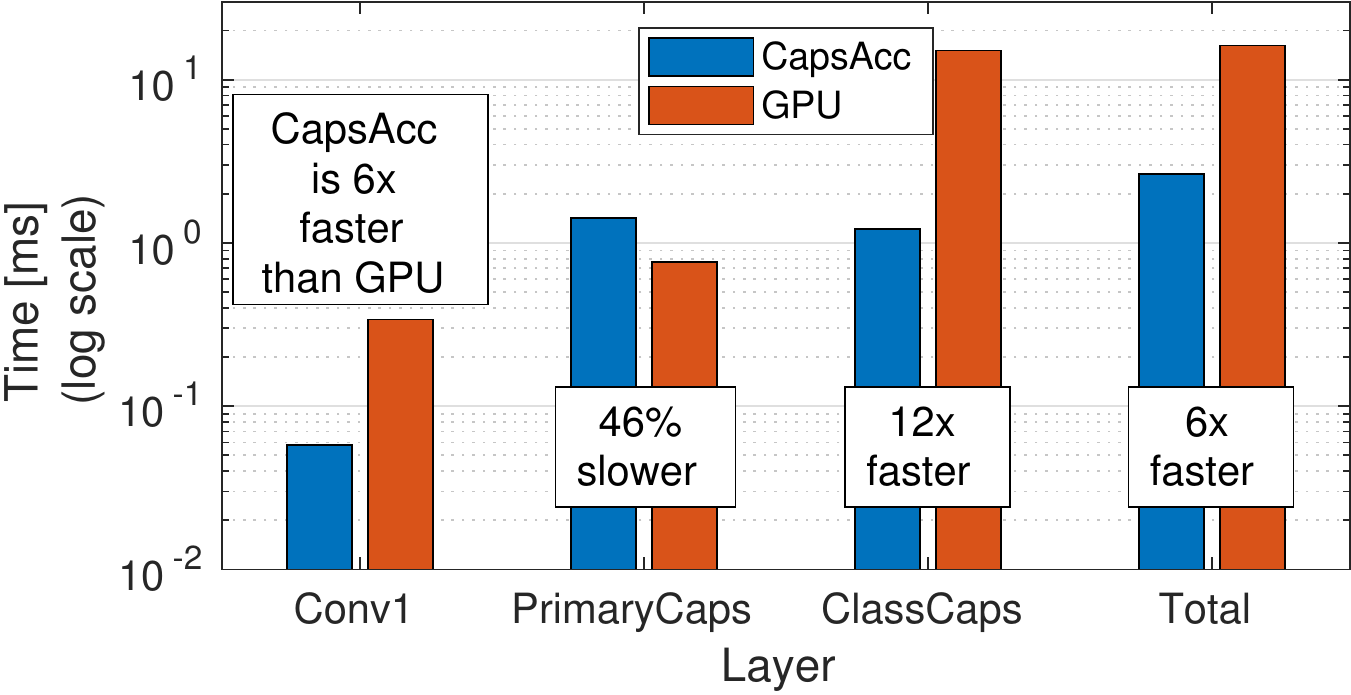}
	\caption{Layer-wise performance of the inference pass on the CapsuleNet on our CapsAcc accelerator, compared to the GPU.}
	\label{fig:capsacc_bar_layers}
\vspace*{0mm}
\end{figure}

\begin{figure}[t]
	\centering
	\includegraphics[width=\linewidth]{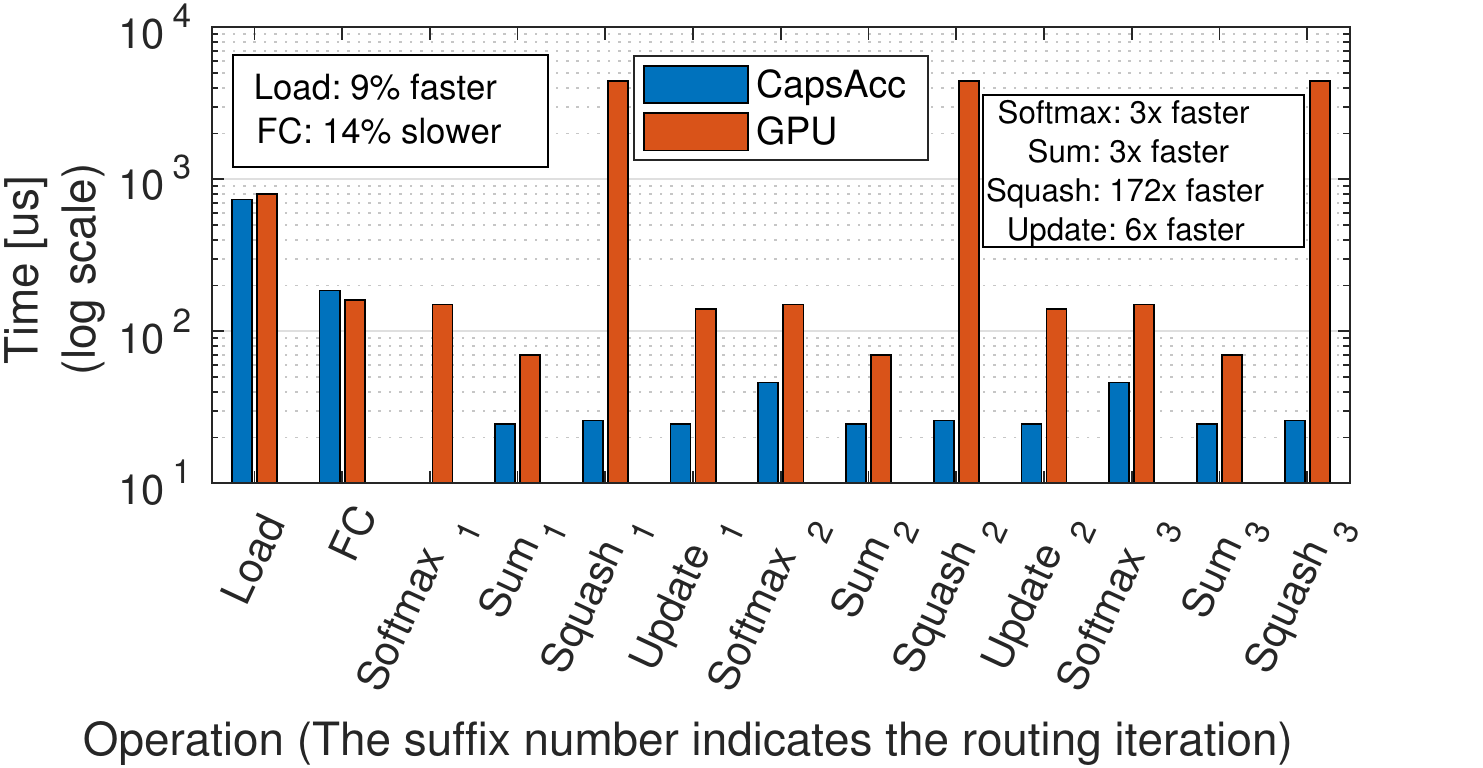}
	\caption{Performance of the inference pass on each step of the routing-by-agreement algorithm on our CapsAcc accelerator, compared to the GPU.}
	\label{fig:capsacc_bar_routing}
\vspace*{0mm}
\end{figure}

The graphs shown in \Cref{fig:capsacc_bar_layers} report the performance (execution time) results of the different layers of CapsuleNet inference on our CapsAcc accelerator, while \Cref{fig:capsacc_bar_routing} shows the performance of every sequence of the routing process. Compared with the GPU performance (see \Cref{fig:bar_layers,fig:bar_routing}), we obtained a significant speed-up for the overall computation time of a CapsuleNet inference pass ($6$$\times$). \textit{The main notable improvements are witnessed in the ClassCaps layer ($12$$\times$) and in the Squashing operation ($172$$\times$).}

\subsection{Detailed Area and Power Breakdown}
\label{subsec:breakdown}

\begin{figure}[t]
\noindent
\begin{minipage}[t]{.40\linewidth}
\vspace{2.6mm}
\begin{table}[H]
\centering
\resizebox{.95\textwidth}{!}{%
\begin{tabular}{|c|c|}
\hline
\textbf{Tech. node {[}nm{]}} & 32 \\ \hline
\textbf{Voltage {[}V{]}} & 1.05 \\ \hline
\textbf{Area {[}mm$^2${]}} & 2.90 \\ \hline
\textbf{Power {[}mW{]}} & 202 \\ \hline
\textbf{Clk Freq. {[}MHz{]}} & 250 \\ \hline
\textbf{Bit width} & 8 \\ \hline
\textbf{On-Chip Mem. {[}MB{]}} & 8 \\ \hline
\end{tabular}
}
\vspace{2mm}
\caption{Parameters of our synthesized CapsAcc accelerator}
\label{tab:arch_parameters}
\end{table}
\end{minipage}
\hspace{.02\linewidth}
\begin{minipage}[t]{.58\linewidth}
\begin{table}[H]
\centering
\resizebox{\textwidth}{!}{%
\begin{tabular}{|c|c|c|}
\hline
\textbf{Component} & \textbf{Area {[}um\textasciicircum{}2{]}} & \textbf{Power {[}mW{]}} \\ \hline
Accumulator & 311961 & 22.80 \\ \hline
Activation & 143045 & 5.94 \\ \hline
Data Buffer & 1332349 & 95.96 \\ \hline
Routing Buffer & 316226 & 22.78 \\ \hline
Weight Buffer & 115643 & 8.34 \\ \hline
Systolic Array & 680525 & 46.09 \\ \hline
Other & 4330 & 0.13 \\ \hline
\end{tabular}
}
\vspace{2mm}
\caption{Area and power, measured for the different components of our CapsAcc accelerator.}
\label{tab:area_power}
\end{table}
\end{minipage}
\vspace*{0mm}
\end{figure}

\begin{figure}[t]
\centering
\begin{minipage}[t]{.34\linewidth}
\vspace*{0mm}
\subfloat[]{
\includegraphics[width=\linewidth]{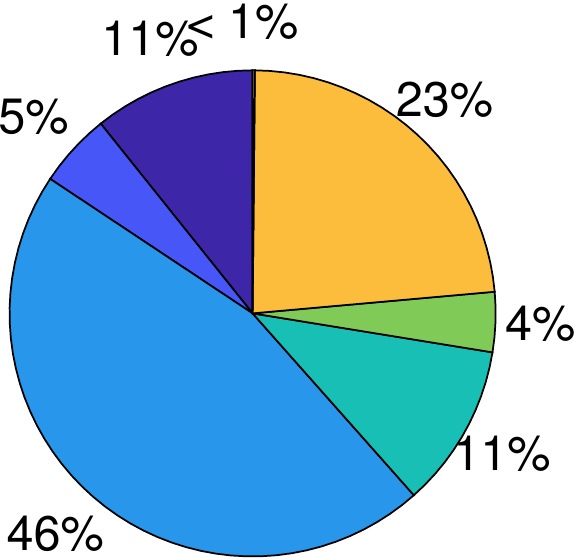}
\label{fig:area_breakdown}}
\end{minipage}
\hfill
\begin{minipage}[t]{.61\linewidth}
\vspace*{0mm}
\subfloat[]{
\includegraphics[width=\linewidth]{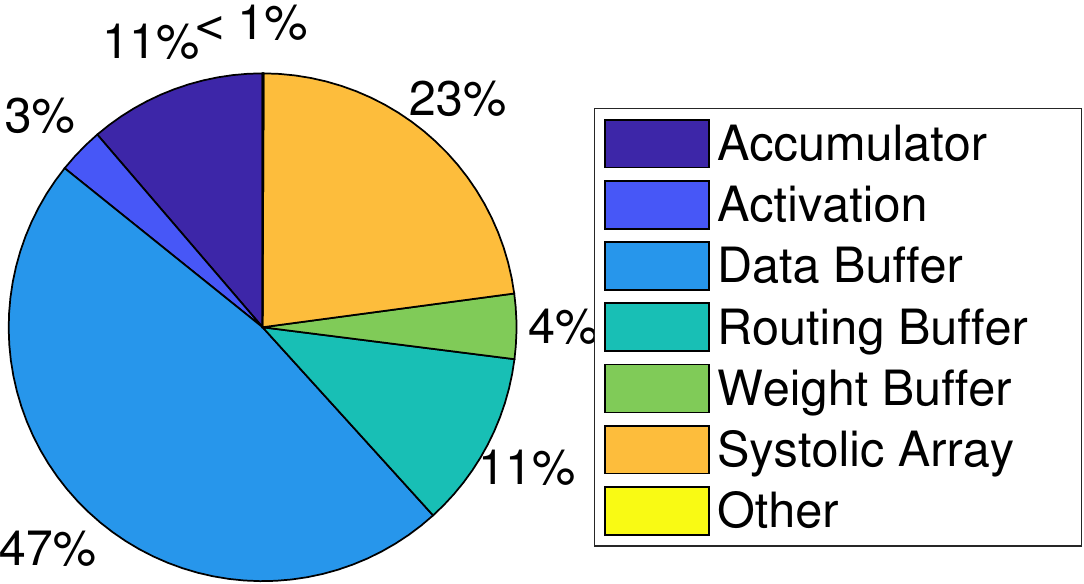}
\label{fig:power_breakdown}}
\end{minipage}
\caption{(a) Area and (b) Power Breakdown of our CapsAcc Accelerator.}
\label{fig:breakdowns}
\vspace*{0mm}
\end{figure}

The details and synthesis parameters for our design are reported in \Cref{tab:arch_parameters}. 
\Cref{tab:area_power} shows the absolute values for the area and power consumption of all the components of the synthesized accelerator. 
\Cref{fig:area_breakdown,fig:power_breakdown} show the area and power breakdowns, respectively, of our CapsAcc architecture. 
These figures show that the area and power contributions are dominated by the buffers, and the systolic array is just 1/4 of the total budget.

\section{Conclusions}
\label{sec:conclusions}

We presented the first CMOS-based hardware accelerator for the complete CapsuleNet inference. To achieve high performance, our CapsAcc architecture employs a flexible systolic array with several optimized data-flow patterns that enable it to fully exploit a high level of parallelism for diverse operations of the CapsuleNet processing. To efficiently use the proposed hardware design, we also optimized the routing-by-agreement algorithm without changing its functionality and thereby preserving the classification accuracy of the original CapsuleNet design of \cite{ref:dyn_routing}. Our results show a significant speedup compared to an optimized GPU implementation. We also presented power and area breakdown of our hardware design. Our CapsAcc provides the first proof-of-concept for realizing CapsuleNet hardware, and opens new avenues for its high-performance inference deployments.

\end{document}